\documentclass[12pt,twoside]{article}
\usepackage[mathscr]{eucal}
\usepackage{amsmath,amsfonts,amssymb,amsthm,ulem}
\usepackage{hyperref}
\usepackage{times}

\usepackage{multibox}

\def\be{\begin{equation}}
\def\ee{\end{equation}}
\def\ba{\begin{array}}
\def\ea{\end{array}}

\def\dps{\displaystyle}

\def\sd{S^\dagger}
\def\bsd{\bar S^\dagger}

\def\BGST{Barnich:2004cr}
\newcommand{\BGadS}{Barnich:2006pc}

\def\Goff{Grigoriev:2006tt}
\def\STV{Shaynkman:2004vu}

\advance\textheight\topskip
\renewcommand{\tilde}{\widetilde}
\renewcommand{\hat}{\widehat}
\newtheorem{prop}{Proposition}[section]

\renewcommand{\simeq}{\cong}

\newcommand{\bref}[1]{\textbf{\ref{#1}}}

\newcommand{\Ker}{\mathop{\mathrm{Ker}}}
\newcommand{\im}{\mathop{\mathrm{Im}}}

\newcommand{\p}[1]{|#1|}
\newcommand{\gh}[1]{\mathrm{gh}(#1)}

\newcommand{\algf}{\Liealg{f}}

\newcommand{\Liealg}{\mathfrak}       %Lie alg[ebra(s)]
\newcommand{\module}[1]{\mathscr{#1}}

\newcommand{\modM}{\module{M}} %%% !!!!!!!!

\newcommand{\dd}{\partial}
\renewcommand{\d}{\partial}

\renewcommand{\geq}{\,{\geqslant}\,}
\renewcommand{\leq}{\,{\leqslant}\,}

\newcommand{\inner}[2]{\langle #1{,}\,#2\rangle}
\newcommand{\binner}[2]{%
  {\langle}\kern-4.15pt{\langle}#1{,}\,#2{\rangle}\kern-4.15pt{\rangle}}
\newcommand{\commut}[2]{[#1{,}\,#2]}

\newcommand{\half}{\mathchoice{%
    \ffrac{1}{2}}{\frac{1}{2}}{\frac{1}{2}}{\frac{1}{2}}}

\newcommand{\ffrac}[2]{\raisebox{.5pt}%
  {\footnotesize$\displaystyle\frac{#1}{#2}$}\kern1pt}

\newcommand{\brst}{\mathsf{\Omega}}

\newcommand{\st}[2]{\overset{#1}{#2}}

\newcommand{\dl}[1]{\mathchoice{\ffrac{\dd}{\dd #1}}{\frac{\dd}{\dd
      #1}}{\ffrac{\dd}{\dd #1}}{\ffrac{\dd}{\dd #1}}}

\newcommand{\vac}{|0\rangle}

\newcommand{\bundle}{\boldsymbol}

\newcommand{\derham}{\boldsymbol{d}}

\newcommand{\manifold}[1]{\mathscr{#1}}
\newcommand{\manX}{\manifold{X}}

\newcommand{\manM}{\manifold{M}}
\def\cE{\mathcal{E}}
\def\cF{\mathcal{F}}
\def\cG{\mathcal{G}}
\def\cH{\mathcal{H}}

\def\cN{\mathcal{N}}

\def\cP{\mathcal{P}}

\def\cS{\mathcal{S}}
\def\cT{\mathcal{T}}

\numberwithin{equation}{section} \makeatletter
\@addtoreset{equation}{section}

\begin{document}

\begin{titlepage}

\begin{flushright}
FIAN/TD/23/08\\
%\texttt{hep-th/yymmnnn}
\end{flushright}

\vspace{0.5cm}
\begin{center}

{\Large\textbf{Massless Poincar\'e modules
\\ [5pt] and gauge invariant equations}}

\vspace{.9cm}

{\large K.~Alkalaev, M.~Grigoriev, and I.~Tipunin}

\vspace{0.5cm}

\textit{I.E. Tamm Department of Theoretical Physics, \\P.N. Lebedev Physical
Institute,\\ Leninsky ave. 53, 119991 Moscow, Russia}

\vspace{1cm}

\begin{abstract}
\noindent Starting with an indecomposable Poincar\'e module $\modM_0$ induced from a given irreducible Lorentz module
we construct a free Poincar\'e invariant gauge theory defined on the Minkowski space. The space of its gauge
inequivalent solutions coincides with (in general, is closely related to) the starting point module $\manM_0$.
We show that for a class of indecomposable Poincar\'e modules the resulting theory is a Lagrangian gauge theory of
the mixed-symmetry higher spin fields. The procedure is based on constructing the parent formulation of the theory.
The Labastida formulation  and the unfolded description of the mixed-symmetry fields are reproduced through
the appropriate reductions of the parent formulation. As an independent check we show that in
the momentum representation the solutions form a unitary irreducible Poincar\'e module determined by
the respective module of the Wigner little group.

\end{abstract}
\end{center}
\end{titlepage}
\newpage

{\small
\tableofcontents
}

\section{Introduction}

Several approaches to massless mixed symmetry fields on the
Minkowski space are known up to now. Although the existence of the
respective irreducible modules in $d>5$ was clear since the famous
Wigner classification~\cite{Wigner:1939cj} finding the covariant
and gauge invariant field equations realizing such modules on
local fields was not completely obvious. Such equations have been
proposed much later by Labastida~\cite{Labastida:1986ft} along
with the candidate Lagrangian \cite{Labastida:1989kw}. The
rigorous proof that the Labastida fields indeed describe
respective representations of the Wigner little group for general
spins was missing till recently \cite{Bekaert:2006ix} (see also
\cite{Bekaert:2006py}). The unfolded form of the equations of
motion and the respective local Lagrangian have been proposed in
\cite{Skvortsov:2008vs, Skvortsov:2008sh}. Massless two-row fields
on Minkowski space have been also analyzed within the BRST
approach in \cite{Burdik:2001hj}. Note also a recently proposed
alternative formulation~\cite{Campoleoni:2008jq} that treats
higher spin fields by relaxing any algebraic
constraints.\footnote{As far as particular cases of mixed-symmetry
fields are concerned there are various successful approaches
available in the literature
\cite{Aulakh:1986cb,Labastida:1986gy,Zinoviev:2003ix}. } The above
approaches have brought to light a number of useful algebraic
structures and field-theoretical methods. However, these
formulations lead to  either quite involved set of covariant
fields and associated algebraic constraints or hidden structure of
the gauge invariance. Moreover, the interrelation between
different approaches and their dynamical equivalence remains
unclear beyond the case of two-row fields.

These problems are mainly due to the lack of unifying  algebraic structures
 underlying the formulations. This calls for the proper
algebraic and dynamical framework that allows one to treat the theory
in model-independent terms and to use the powerful machinery of the
representation theory combined with an effective technique to handle the involved gauge symmetry and the constraints present in the
models.

In this paper we take a rather abstract point of view and describe
a class of massless Poincar\'e modules in terms of Howe dual pair of
Lie algebras: the Lorentz algebra $o(1,d-1)$ and symplectic algebra $sp(2n)$
represented on the suitable polynomials ($n-1$ corresponds to the
number of rows in the Young tableau of the respective covariant
field). It turns out that the massless Poincar\'e modules that can be
realized on local fields naturally arise as quotient spaces rather
than just subspaces of polynomials. This is crucial because in the
field theory this quotient construction is realized through the
gauge invariance.

Another important  ingredient is the BRST (cohomology) technique
that allows one to translate the pure algebraic definition of the
Poincar\'e module into the genuine local gauge field theory. This
can be seen as a far going generalization of the following
procedure known in the literature (see,
{\it e.g.},~\cite{Shaynkman:2004vu} for the discussion in the related
context): given an $\algf$-module $\modM_0$ and a manifold $\manX$
equipped with a flat $\algf$-connection one considers
$\modM_0$-valued field subjected to the covariant constancy
condition understood as an equation of motion. The space of
solutions to this equation coincides with (in general, is closely
related to) $\modM_0$ and the system is explicitly invariant under
$\algf$. However, this construction does not directly lead to
gauge invariant equations. In particular, this makes the equations
of motion in general non-Lagrangian. Moreover, studying possible
interactions becomes complicated because nonlinear deformations
are usually formulated in terms of gauge potentials.

The procedure proposed in this paper allows one to find a complete
set of gauge fields needed for the gauge theory description of the
given Poincar\'e module $\modM_0$.\footnote{From this perspective
our approach can be viewed as somewhat similar to the method of
covariantized light-cone developed in \cite{Siegel:1986zi}.} More
precisely we consider the indecomposable Poincar\'e modules induced
from a given irreducible Lorentz module determined by spins
$s_{n-1}\geq s_{n-2}\geq    \ldots \geq s_1$. The idea is to
realize the Poincar\'e module $\modM_0$ as the ghost-number-zero
cohomology of the appropriate BRST operator $Q$. Using the BRST
extension~\cite{\BGST,\BGadS,\Goff} of the unfolded
formalism~\cite{Vasiliev:1988xc,Vasiliev:1988sa,Shaynkman:2000ts,Vasiliev:2001wa,Bekaert:2005vh,Skvortsov:2008vs}
allows us to immediately construct the local gauge field theory by
replacing the covariant constancy condition with its BRST
extension using the generalized covariant derivative
$\hat\brst=\nabla +Q$. This derivative is naturally interpreted as
a BRST operator of a first-quantized constrained system so that
the constructed field theory is a free field theory associated to
this quantum constrained system. In the case $n=2$ (totally symmetric fields)
this formulation was identified in~\cite{\BGST}.

It has to be stressed that using the BRST technique brings in the
ghost grading that selects physical fields (those at ghost number
zero) among all the fields entering the BRST extended formulation.
It turns out, that besides the $\modM_0$-valued fields one finds
other ghost number zero fields that are necessarily differential
forms of nonvanishing degree. These fields are automatically gauge
fields, with the gauge transformations and the reducibility
relations determined by the BRST operator $\hat\brst$.

Using the method developed in~\cite{\BGST,\BGadS} allows us to
extend the formulation based on $\hat\brst$ to an equivalent
formulation where some of the algebraic constraints are
implemented implicitly by the appropriately extended BRST
operator. This determines a proper counterpart of the parent
theory from~\cite{\BGST} that serves to obtain various other
formulations through the equivalent reductions (elimination of
generalized auxiliary fields). In particular, we show that the
parent theory reduces to the well-known Labastida
theory~\cite{Labastida:1986ft,Labastida:1989kw} and the recently constructed
unfolded formulation~\cite{Skvortsov:2008vs}. As a byproduct this
gives a proof that these two formulations are locally equivalent
at the level of equations of motion,
{\it i.e.} the equivalence of the metric-like and the frame-like local formulations.

Making use of the parent theory allows one to find another particular
reduced theory that admits a standard Lagrangian of the form
$\langle\Psi,\Omega \Psi\rangle$. This has the same structure as
the analogous Lagrangian for Fronsdal HS fields proposed in
\cite{Bengtsson:1986ys,Ouvry:1986dv,Henneaux:1987cpbis}.
Just BRST operator $\Omega$ entering the action is known in the literature
as an appropriate truncation of the open bosonic string BRST operator
in the tensionless limit~\cite{Sagnotti:2003qa,Bonelli:2003kh}.
What we prove here is that this Lagrangian indeed describes an irreducible
mixed-symmetry field provided the appropriate set of algebraic constraints are
imposed on $\Psi$.

As an
independent check we show that the space of gauge inequivalent
solutions of the model in the momentum representation with $p\neq
0$ indeed coincides with the irreducible unitary module induced
from the respective module of the Wigner little group with the
same spins $s_{n-1}\geq s_{n-2}\geq    \ldots \geq s_1$. This
shows that the constructed theory indeed describes the unitary dynamics
of the right number of physical degrees of freedom.

The approach developed in the paper can be applied far beyond the
context of the Poincar\'e invariant equations. In particular, it can
be extended to cover the linear equations in the $AdS_d$ space, where
the parent formulation is known~\cite{\BGadS} for the case of
Fronsdal fields, {\it i.e.} $n=2$. There also remains to see how the
massive fields can be described in this way. More precisely, how
the dimensional reduction can be implemented in these terms. More
ambitious perspective has to do with describing the gauge field
realization of $F$-modules (for $F$ sufficiently general) on the
homogeneous spaces $F/G$.

The paper is organized as follows: the main construction is
presented in detail in Section~\bref{sec:brst}. There  we also
introduce most of the technical tools needed throughout the paper.
In Section~\bref{sec:Labastida} we show how the field content and the
equations of motion of the Labastida theory can be obtained by an
appropriate reduction of the parent formulation and discuss its relation
to the tensionless limit of string theory.
Section~\bref{sec:GPM} is devoted to the analysis of various
Poincar\'e modules appearing in the different formulations. This
involves explicit reduction to the unfolded form and establishing
a relationship with the modules of the Wigner approach.
Conclusions and perspectives
are discussed in Section~ \bref{sec:conc}.

\section{BRST operator for mixed-symmetry fields on Minkowski space}\label{sec:brst}

\subsection{Howe dual realization of the Poincar\'e algebra}
\label{sec: howe}

Let us start with Minkowski space $ISO(1,d-1)/SO(1,d-1)$ whose algebra of infinitesimal isometries
 is the Poincar\'e algebra $iso(1,d-1)$. We denote the basis elements
 of the Poincar\'e algebra as $P_a$ and $M_{ab}$ (translations and Lorentz transformations).
Suppose we are interested in the representations induced from the finite-dimensional
irreducible representations of the Lorentz subalgebra $so(1,d-1)$.
It is useful to discuss first the subspaces irreducible under the Lorentz subalgebra
that can be nicely described using the following oscillator realization.

Let us introduce bosonic variables $a^a_I$ and $\bar a^J_b$, $\;\;a,b=0, ..., d-1$, $\;I,J=0, ...,n-1$ satisfying the
canonical commutation relations
\begin{equation}
\label{oscillatros}
[{\bar a}_a^I, a^b_{J}] = \delta^I_J\delta_a^b\,.
\end{equation}
It is assumed that $\bar a^I_a$ acts as $\dps\dl{a_I^a}$ on the space $\cP^d_{n}(a)$ of polynomials in $a_I^a$
\be
\label{pol}
\phi(a) = \sum_{m_I}\phi_{a_1\;\ldots\; a_{m_0};\;\ldots\ldots\;;\, c_1 \;\ldots\; c_{m_{n-1}}} a^{a_1}_0 \cdots a^{a_{m_0}}_0
\;\cdots\;
a^{c_1}_{n-1} \cdots a^{c_{m_{n-1}}}_{n-1}\;,
\ee
where $m_I\equiv (m_0, ..., m_{n-1})$ are arbitrary non-negative integers.
Introducing the Minkowski metric $\eta_{ab}$ one can represent the Lorentz algebra
on $\cP^d_{n}(a)$ as
\begin{equation}
\label{lorgen}
 M_{ab}=a_I{}_a\bar a^{I}_b-a_I{}_b\bar a^{I}_a\,.
\end{equation}
Here and in what follows indices  $a,b$ are raised and lowered using the Minkowski
metric. It follows that the expansion coefficients in (\ref{pol}) transform
as Lorentz tensors. The space of all polynomials decomposes into the finite-dimensional
irreducible modules of the Lorentz algebra.
In order to describe all the finite-dimensional modules with
integer spins in a given dimension $d$ one needs to take $n=[\frac{d}{2}]$.

It is useful to study the structure of $\cP^d_{n}(a)$ as the module over the orthogonal algebra
$so(1,d-1)$ using the
Howe duality~\cite{Howe, Howe1}. The  Howe dual algebra to the $so(1,d-1)$ algebra is
$sp(2n)$ algebra  with the basis elements given by \cite{Howe, Howe1}
\begin{equation}
\label{SPgenerators}
 T_{IJ}=a_I^a a_{Ja}\,,\quad T_I{}^J=\frac{1}{2}\,(a^a_I \bar a^J_a+\bar a^J_a a^a_I)\,, \quad T^{IJ}=\bar a^I_a\bar a^{Ja}\,.
\end{equation}
Their non-zero commutation relations read
\be
\nonumber
\ba{c}
\dps
[T_I{}^J, T_K{}^L]= \delta_K^J T_I{}^L-\delta_I^LT_K{}^J,
\quad
[T^{IJ}, T_{KL}] = \delta^I_K T_L{}^J+\delta^I_L T_K{}^J+\delta^J_K T_L{}^I+\delta^J_L T_K{}^I,
\\
\\
\dps
[T_K{}^L, T_{IJ}]= \delta_J^L T_{KI}+\delta^L_IT_{KJ} \;,
\quad
[T^{IJ}, T_K{}^L]=\delta^I_K T^{JL}+ \delta^J_KT^{IL} \;.
\ea
\ee
The diagonal elements $T_I{}^I$ form a basis in the Cartan subalgebra while $T^{IJ}$ and $T_I{}^J, I>J$ are the basis elements
of the upper-triangular subalgebra. Let us note that $gl(n)$ algebra is realized
by the generators
$T_I{}^J$ as a subalgebra of $sp(2n)$ while its $sl(n)$ subalgebra is generated by $T_I{}^J$ with $I\neq J$.

The finite-dimensional  irreducible representations of the Lorentz algebra in the space of
polynomials in $a^a_I$ are singled out by the
highest weight conditions of the dual $sp(2n)$, {\it i.e.} annihilated by the upper triangular subalgebra of $sp(2n)$ along
with the weight conditions with respect to the Cartan subalgebra.
In addition, to describe all integer spin  finite-dimensional Lorentz irreps one needs to take
$n\leq \nu$, where $\nu=[\frac{d}{2}]$ is a rank of the Lorentz algebra $so(1,d-1)$.
More precisely, let $s_I$ be integer numbers such that $s_I\geq s_J$ for $I>J$.
We assume that the following weight conditions are imposed
\begin{equation}
\label{weight}
 T_I{}^I\phi=(s_I+\frac{d}{2})\,\phi\,.
\end{equation}
Imposing then the tracelessness and Young symmetry conditions
\begin{equation}
\label{upp-triang-cond}
 T^{IJ}\phi=0\,, \qquad T_I{}^J\phi=0\,\quad I>J\,,
\end{equation}
one gets a finite-dimensional irreducible representation of the Lorentz algebra described
by Young tableau of the symmetry type $(s_{n-1}, s_{n-2}, \cdots, s_0)$
\be
\label{YT1}
\begin{picture}(75,70)(0,2)
\multiframe(0,0)(10.5,0){1}(10,10){}\put(20,0){$s_0$}
\multiframe(0,10.5)(10.5,0){2}(10,10){}{}\put(25,12.5){$s_{1}$}
\put(-0.2,20){\line(0,1){25}}
\put(9,26){$\vdots$}
\multiframe(0,41.5)(10.5,0){5}(10,10){}{}{}{}{} \put(60,42.5){$s_{n-2}$}
\multiframe(0,52)(10.5,0){7}(10,10){}{}{}{}{}{}{} \put(79,55){$s_{n-1}$}
\end{picture}
\ee

\vspace{.2cm}

Let us now briefly recall the formal structure of the polynomials in $a^a_I$ as a module
over the Howe dual $so(1,d-1)$ and $sp(2n)$ algebras. More detailed discussion can be found
in the Appendix~\bref{sec:A}, where we also collect some useful statements needed in the main text.
$\cP^d_n(a)$ considered as a $so(1,d-1)$ and $sp(2n)$ bimodule can be lifted to the respective complex module of the
complexified algebras. The structure of the irreducible components is unchanged under the
complexification. This allows us to use the results known in the literature.
Since $so(1,d-1)$ and $sp(2n)$ algebras obviously commute,
the space of polynomials $\cP_{n}^d(a)$
is a $so(d)$ -- $sp(2n)$ bimodule.
For $n\leq [\frac{d}{2}]$ bimodule $\cP_{n}^d(a)$
has the following structure \cite{AdBarb}
\be
\label{Pnd}
\cP_{n}^d = \mathop{\oplus}\limits_{\sigma\in\Lambda}
(V_\sigma \otimes U_{\theta(\sigma)})\;,
\ee
where $V_\sigma$ and $U_{\theta(\sigma)}$ are respectively irreducible $so(d)$ and $sp(2n)$
modules with highest weights $\sigma$ and $\theta(\sigma)$, where $\theta$ is some mapping
(for more details see Appendix~\bref{sec:A}).
While $V_{\sigma}$ is finite-dimensional $U_{\theta(\sigma)}$ is the generalized Verma module
induced from the finite-dimensional irreducible $sl(n)$ module (more precisely,
from the module of the corresponding parabolic subalgebra in $sp(2n)$).
In particular, this implies that $U_{\theta(\sigma)}$ is freely
generated by generators $T_{IJ}$ from the respective $sl(n)$-module~\footnote{That is besides the $sp(2n)$ algebra relations there
are no additional relations between elements of the form $T_{I_1J_1}T_{I_2J_2}\ldots T_{I_kJ_k}\phi$ with $\phi$ in $sl(n)$-module. For instance,
as a linear space $U_{\theta(\sigma)}$ is isomorphic to polynomials in $T_{IJ}$ with coefficients in the $sl(n)$-module.}.

\subsection{Poincar\'e modules}

Remarkably the set of oscillators (\ref{oscillatros}) allows one to realize
the Poincar\'e algebra as well. To this end we  relax some of the conditions \eqref{weight} and \eqref{upp-triang-cond} in order to
describe some infinite-dimensional (indecomposable) representations of the Poincar\'e algebra.
First of all we choose the Poincar\'e
generators $P_a$ to act as ``translations'' for the $I$-th oscillators. Without loss of generality
we take $I=0$ so that~\footnote{Inequivalent but somehow dual choice is to take
$P_a=a^0_a$. This would lead to an indecomposable representation freely generated
from a given Lorentz representation in contrast to the co-freely generated one which
we are going to get.}
\be
\label{Poinctrans}
\dps
P_a=\dl{a^a_0}\;.
\ee
In the sequel  we use the following notations
$a_0\equiv y,\;a_I\equiv a_i\,,\,I>0$ with $i=1,..., n-1$.
Furthermore, it is convenient to introduce special
notations for some $sp(2n)$ generators
\begin{equation}
\label{newnot}
\begin{gathered}
 \sd_i\equiv T_i{}^0=a_i^a\dl{y^a}\,,\qquad \bsd{}^i\equiv T_0{}^i=y^a\dl{a_i^a}\,,\\
 N_i{}^j\equiv T_i{}^j=a_i^a\dl{a_j^a}\,\,\, i\neq j\,,\qquad
N_i\equiv T_i{}^i-\frac{d}{2}=a_i^a\dl{a_i^a}\,.
\end{gathered}
\end{equation}

Operators $P_a$ obviously commute with all the irreducibility conditions but $T_0{}^0\phi=(s_0+\frac{d}{2})\phi$.
By relaxing this condition one gets a representation (in fact indecomposable) of the Poincar\'e algebra.
This representation is finite-dimensional as the conditions
$\sd_i\phi=0$ imply that for a homogeneous element a homogeneity degree in $y$ is
lower than that in $a^a_i$. In other words, the operator $P_a$ acts on the last row of
the corresponding Young tableau by shortening its length and the whole carrier space
consists of Lorentz irreps described by Young tableaux
(\ref{YT1}) with $0\leq s_0\leq s_1$. To summarize, the resulting Poincar\'e module
is singled out by the conditions
\begin{gather}
\label{Gauge-module}
 T^{IJ}\phi=0 \,,\qquad N_i\phi=s_i\,\phi \,,\qquad N_i{}^j\phi=0\;\;\;\;\,\,\, i>j\,, \\
 \sd_i\phi=0\,.
\label{Gauge-module-2}
\end{gather}

Although the Poincar\'e module just constructed plays an important
role in the subsequent analysis it is not the one we are
interested in now. This is because a representation realized on
local fields is necessarily infinite-dimensional. In order to
arrive at an infinite-dimensional module let us consider a
subspace $\modM_0$ singled out by a slight modification of \eqref{weight}
and \eqref{upp-triang-cond}. Namely, in addition to relaxing
$T_0{}^0\phi=(s_0+\frac{d}{2})\phi$ we also invert the Young
conditions involving $y^a\equiv a_0^a$ so that the full set of the
conditions reads explicitly as
\begin{gather}
\label{Weyl-module}
 T^{IJ}\phi=0 \,,\qquad N_i\phi=s_i\,\phi \,,\qquad N_i{}^j\phi=0\;\;\;\;\,\,\, i>j\,, \\
 \bar S^\dagger{}^i\phi=0\,.
\label{Weyl-module-2}
\end{gather}
%where we use notation $\displaystyle(\bar\cS^\dagger)^i\equiv T_0{}^i=y^a\dl{a^a_i}$.
For the corresponding Young tableau
this implies a rearranging the rows by moving the last row to the top as expressed by the last condition.
The resulting module $\modM_0$ is described by an infinite  collection of Young tableaux
\be
%\label{YT2}
\begin{picture}(75,70)(0,2)
\multiframe(0,0)(10.5,0){1}(10,10){}\put(20,0){$s_1$}
\multiframe(0,10.5)(10.5,0){2}(10,10){}{}\put(25,12.5){$s_{2}$}
\put(-0.2,20){\line(0,1){25}}
\put(9,26){$\vdots$}
\multiframe(0,41.5)(10.5,0){5}(10,10){}{}{}{}{} \put(60,42.5){$s_{n-1}$}
\multiframe(0,52)(10.5,0){7}(10,10){}{}{}{}{}{}{} \put(79,55){$s_0$}
\end{picture}
\ee

\vspace{.2cm}

\noindent with running $s_0$ bounded from below, $s_0\geq s_{n-1}$. Although the condition in
(\ref{Weyl-module-2}) does not commute with $P_a$ one can consistently define the action of $P_a$ on the
subspace using the appropriate projector. The quadratic Casimir operator of the Poincar\'e
algebra $C_2=P^2=T^{00}$ is automatically zero on module $\modM_0$ because of \eqref{Weyl-module}
so that $\modM_0$ is the massless Poincar\'e module. In the unfolded description of Fronsdal fields on $AdS_d$ the respective counterpart of
$\modM_0$ is often referred to as Weyl module.

The origin of the difference between the Poincar\'e module determined by \eqref{Gauge-module},\eqref{Gauge-module-2}
and $\modM_0$ is that they are described  by the highest weight
conditions with respect to the two different choices of the upper triangular
subalgebra of $sl(n)\subset sp(2n)$. They are generated by
($N_i{}^j\;\; i>j$, $S_i^\dagger$) and ($N_i{}^j\;\; i>j$,
$\bsd{}^i$), respectively. Moreover, the irreducibility conditions for these
Poincar\'e modules contain the subalgebra formed by $N_i{}^j\,\,\, i>j$.
In fact there are other choices for the upper triangular subalgebra
containing  $N_i{}^j\,\,\, i>j$ that play the essential role in the subsequent analysis.

\subsection{BRST realization}\label{sec:Q}
It turns out that subspace $\modM_0$ defined by \eqref{Weyl-module} and \eqref{Weyl-module-2} can be represented
in an explicitly Poincar\'e invariant way. The idea is to identify it as an appropriate quotient of a Poincar\'e
invariant subspace with respect to a Poincar\'e invariant equivalence relation.
Indeed, as we have noted $\modM_0$ is defined by the highest weight conditions (for an appropriate
choice of weight ordering)
of the $sl(n)$ algebra generated by  $N_i{}^j\,\,\, i> j$ along with
$\bsd{}^i$ and $\sd_i$. By decomposing the entire space into the finite-dimensional irreducible $sl(n)$ components
one finds that in each component the only element satisfying
$N_i{}^j\phi=0\,\,\, i> j$ and not in the image of any of $\sd_i$ is the highest
weight vector $\bsd{}^i\phi=0$.
Because generators $\sd_i$ obviously commute with $P_a$ we arrive at the
following Poincar\'e invariant equivalence relation
\be
\label{eqrel}
\phi(y,a) \sim \phi(y,a) + S^\dagger_i \phi^i(y,a)\;,
\ee
and hence the representatives can be identified with those satisfying \eqref{Weyl-module-2}.

It appears useful to implement this construction in the BRST terms. To this end we introduce fermionic
ghost variables $\bar c^i,\, b_i,\; i=1,...,n-1$ satisfying~\footnote{Here and in what follows
the commutator denotes the graded commutator, $\commut{f}{g}=fg-(-)^{|f||g|}gf$,
where $|f|$ is the Grassmann parity of $f$.}
\begin{equation}
 \commut{\bar c^i}{b_j}=\delta^i_j\,,\qquad \gh{\bar c^i}=1\,,\quad \gh{b_i}=-1\,,
\end{equation}
where $\gh{\cdot}$ denotes the ghost degree.
These variables are represented on functions of $b^i$ as $\dps\bar c^i\phi=\dl{b_i}\phi$.
We consider the following BRST operator
\begin{equation}
 Q=\bar c^i S^\dagger_i\,.
\end{equation}
Because the constraints form the Abelian algebra $[S^\dagger_i, S^\dagger_j]=0$
the terms cubic in ghosts are absent in the BRST operator.

The space \eqref{Weyl-module}
can be identified then with the ghost-number-zero cohomology of $Q$
evaluated in the space of elements satisfying
\begin{equation}
%\begin{gathered}
\label{part-Weyl}
\hat\cN_i\phi\equiv (N_i+b_i\bar c^i)\phi=s_i\,\phi\,,\ \ \
\hat\cN_i{}^j\phi\equiv (N_i{}^j+b_i \bar c^j)\phi=0\,\,\, i>j\,, \ \ \
T^{IJ}\phi=0\,.
 %\end{gathered}
\end{equation}
Here operators $\hat\cN_i$ and $\hat\cN_i{}^j$ are the BRST invariant extensions of the respective operators
in \eqref{Weyl-module}, {\it i.e.} $\commut{Q}{\hat\cN_i}=\commut{Q}{\hat\cN_i{}^j}=0$. As for the trace operators
 $T^{IJ}$ they are imposed directly because all the remaining conditions
and the BRST operator $Q$ preserve the subspace singled out by $T^{IJ}\phi=0$. It is easy to check that conditions
\eqref{part-Weyl} are consistent and $Q$ acts in subspace \eqref{part-Weyl}.
Moreover, in the zeroth ghost degree ({\it i.e.} $b_i$-independent elements) conditions \eqref{part-Weyl} explicitly coincides with
the conditions~\eqref{Weyl-module}.

A useful way to
see that the construction is consistent is to observe
that all the constraints \eqref{part-Weyl} along with the constraint $\sd_i$
entering the BRST operator form the upper-triangular subalgebra of $sp(2n)$
completed by the weight conditions from the diagonal (Cartan) subalgebra.
An alternative way to implement the construction is to impose all these
constraints  by the appropriate BRST operator and require in addition the cohomology representatives
to be independent of all the ghost variables but $b_i$. In fact a similar representation is going
to be useful in Sections~\bref{sec:standard} and \bref{sec:Q-coh}.

Because translations $P_a$ and Lorentz generators $M_{ab}$ obviously commute with $Q$
and conditions \eqref{part-Weyl} the Poincar\'e algebra acts in the cohomology.
At the same time, the
zero-ghost-number cohomology is given by the ghost-independent
elements quotient over the image of
$S_i^\dagger$ leading to equivalence relation \eqref{eqrel}.
To see that representatives of these equivalence classes can be
chosen to satisfy $\bsd{}^i \phi=0$ we note that for a ghost-independent element
conditions $\hat\cN_i{}^j\phi=0\,\,\, i>j$ reduce to $N_i{}^j\phi=0\,\,\, i>j$.
This shows that the zero-ghost-number $Q$-cohomology indeed
coincides with module $\modM_0$.

Remarkably, $Q$-cohomology  in other ghost
degrees is in general nonempty. It is  represented by the highest weight vectors
for other choices of the upper triangular subalgebra of $sl(n)$ containing $N_i{}^j\,\,\, i>j$.
A detailed discussion  will be given in Section \bref{sec:Q-coh}.

\subsection{Poincar\'e module of the solutions to PDE on Minkowski space}
We now address a question of how a Poincar\'e module can be realized on the space of
solutions of a system of differential equations on Minkowski space. This can be achieved using
the construction known in the literature (see {\it e.g.} \cite{\STV} for the discussion in the related context). The construction can be
formulated in rather general terms. Namely, let $\modM_0$ be an $F$-module ($F$ being a Lie group, not
necessarily the Poincar\'e group).  Let also $\manX=F/G$ with $G\subset F$ be a symmetric space so that
there is a canonical principle $F$-bundle over $\manX$. One then constructs the associated
vector bundle with the fiber being $\modM_0$. There is a flat $\algf$-connection
(originating from the canonical $\algf$-valued form on $F$; here $\algf$ is a Lie algebra of $F$)
on the principle $F$-bundle over $\manX$, which
determines a flat connection $\alpha$ in the associated vector bundle.

Using the $\algf$-connection $\alpha$ one can represent $\modM_0$
as the space of covariantly constant sections of the associated vector bundle,
{\it i.e.} sections satisfying
\begin{equation}
\label{covconst}
\nabla \Phi=0\,, \qquad \nabla=dx^a(\dl{x^a}+\alpha_a)\,,
%\,\,\,\text{in the standard frame}\,\,)\,.
\end{equation}
where $x^a$ are local coordinates on $\manX$.
Indeed, the space of solutions to this equation in the appropriate functional space is isomorphic to
(in general, closely related to) the fiber at a given point, {\it i.e.} $F$-module $\modM_0$.

Let us discuss how the Lie algebra $\algf$ of $F$ acts on solutions. To this end let $L_A$ be a basis in $\algf$
and by a slight abuse of notations we also denote by $L_A$ the action of $L_A$ in $\modM_0$.
As usual in the field theory it is useful to define the action on fields such that the base space $\manX$ is not affected.
Let in a given point $p\in\manX$ the algebra acts on the field according to
$\lambda|_{p} \Phi|_p=(\lambda^A|_p) L_A \Phi|_p$, where $\lambda^A|_p$ are components of $\lambda$. This action can be
uniquely extended on $\manX$ to the action of the
form $\lambda^A(x) L_A \Phi(x)$
by requiring
\begin{equation}
\label{action-def}
 \commut{\nabla}{\lambda^A(x) L_A}=d\lambda(x)+\commut{\alpha^A L_A}{\lambda^B(x) L_B}=0\,,
\end{equation}
{\it i.e.} the action on the field is determined by a covariantly constant section $\lambda^A(x)$ of the associated vector bundle with the fiber
being $\algf$. This guaranties that \eqref{covconst} is indeed $F$-invariant.

This construction is easily specialized to the case where $F$ is a Poincar\'e group, $G$ its
Lorentz subgroup, $\manX$ Minkowski space, and $\modM_0$ the Poincar\'e module considered above,
and $P_a,M_{ab}$ are the Poincar\'e generators in $\manM_0$. In the Cartesian coordinates $x^a$ on $\manX$ the connection form $\alpha$ can be chosen to be $\alpha=-dx^a P_a$ so that \eqref{covconst} takes the form
\begin{equation}
\label{covconst-P}
dx^a(\dl{x^a}-P_a)\Phi=0\,.
\end{equation}
The action of the Poincar\'e generators on fields (sections) can be obtained from \eqref{action-def}. Namely, the translations and Lorentz rotations act respectively as
\begin{equation}
\label{hat-Poincare}
 \hat P_a \Phi=P_a\Phi\,,\quad \hat M_{ab}\Phi=M_{ab}\Phi+x_a P_b\Phi-x_b P_a\Phi\,.
\end{equation}
Modified generators $\hat P,\hat M$ satisfy the same algebra. Recall that $P_a$ denotes an appropriate projection of  $\dl{y^a}$ on module $\modM_0$.

The above construction can be illustrated in the case
$n=1$ where variables $a^a_i$ are not present\,\footnote{This case corresponds to the Klein--Gordon field. The respective equations of motions in the form \eqref{covconst-P} were thoroughly studied in \cite{Shaynkman:2000ts}.}
so that the Poincar\'e generators in $\modM_0$
do not require projectors and are given by $P_a=\dl{y^a}$ and $M_{ab}=y_a\dl{y^b}-y_b\dl{y^a}$.
Their $x$-dependent realizations ({\it i.e.} action on $\modM_0$-valued sections) read as
\be
 \hat P_a \Phi=\dl{y^a}\Phi\,,\quad \hat M_{ab}\Phi=(x_a+y_a) \dl{y^b}\Phi-(x_b+y_b)\dl{y^a}\Phi\,.
\ee
In this simple example the covariant constancy condition \eqref{covconst-P} just says that
$\Phi(x,y) = \Phi(x+y,0)=\Phi(0,x+y)$.

The Poincar\'e invariant equations \eqref{covconst-P} are not
completely satisfactory from various viewpoints. First of all,
there is no gauge symmetry. More precisely, as we are going to see
$\modM_0$-valued fields can be identified with gauge-invariant HS
curvatures. What is more important, equations \eqref{covconst-P} are
not likely to be Lagrangian even if one adds/eliminates auxiliary
fields (recall that already Maxwell equations are Lagrangian only
if one introduces potentials and hence the gauge symmetry). In
addition, the Poincar\'e algebra is in general realized by the
operators involving projectors in contrast to the realization on
polynomials or their Poincar\'e invariant subspaces.

Before replacing \eqref{covconst-P} with a genuine gauge theory let
us also note that strictly speaking, as solutions to the equation
\eqref{covconst-P} one only gets polynomials
in $x^a$ because in the fiber we have not allowed for elements
non-polynomial in $y^a\equiv a_0^a$. The way out is to consider a
somehow maximal fiber~\footnote{This choice is natural from the first-quantized point of view (see \cite{Fedosov:1996fu}).}  that is the space of elements that are
formal power series in $y^a$ and polynomials in the
remaining oscillators. In this way one can describe solutions from,
\textit{e.g.}, $C^\infty(\manX)$. Note, however, that in this setting the space of
solutions is not isomorphic to the fiber because there can be nonconvergent power series
that cannot be extended to a smooth covariantly constant sections.
In what follows we assume formal power series in $y^a$ variables.

\subsection{Intermediate formulation}

In order to be able to obtain genuine gauge symmetries in this
framework we are going to replace the Poincar\'e module $\modM_0$ with
a graded Poincar\'e module $\modM$ containing $\modM_0$ at zeroth degree and
then consider a gauge theory associated to this graded space in a
similar way as non-gauge theory \eqref{covconst-P} is associated to
$\modM_0$. In fact, we already have all the requisites for this
generalization. Indeed, the cohomology of $Q$ evaluated in
\eqref{part-Weyl} is a Poincar\'e module graded by the ghost degree
such that $\modM_0$ is its degree zero subspace. Moreover, it is
well known how the construction \eqref{covconst} can be
generalized once the module is described in terms of the BRST
operator. This generalization is known as a BRST extended unfolded
formulation. It has been proposed in~\cite{\BGST,\BGadS} (see also
\cite{\Goff}) in constructing the so-called parent formulations of
the linear gauge theories.

The construction of the BRST extended unfolded formulation proceeds as follows. Replacing $dx^a$ with
the Grassmann odd ghost variables $\theta^a, \gh{\theta^a}=1$ one extends the BRST operator $Q$ to
\begin{equation}\label{parent0}
\hat\brst=\nabla+Q\,,\qquad \nabla=\theta^a(\dl{x^a}-\dl{y^a})\,,
\end{equation}
and takes as a representation space functions in $x^a$ with values in
 the tensor product $\hat\cH$ of the representation space for $Q$
 ({\it i.e.}, the space of formal series in $y^a$ and polynomials in $a^a_i,b_i$ satisfying \eqref{part-Weyl})
and the Grassmann algebra generated by $\theta^a$.
Although the theory \eqref{parent0} is explicitly written in Cartesian
coordinates on $\manX$ and the adapted local frame it can easily be rewritten
in terms of arbitrary coordinates $x^\mu$ and
arbitrary local frame using a more general flat covariant derivative
$\nabla=\theta^\mu(\dl{x^\mu}-e_\mu^a\dl{y^a}-\omega_{\mu a}^b (y^a \dl{y^b}+a_i^a\dl{a_i^b}))$. Here $e_\mu^a$ and $\omega_{\mu a}^b$ are coefficients
of the flat Poincare connection $\alpha$ and are to be identified with the vielbein and the Lorentz connection on $\manM_0$.

Given a BRST operator $\hat\brst$ represented on $\hat\cH$-valued functions in $x^a$
the associated gauge field theory is determined by
the BRST differential $s$ defined through $s\Psi=\hat\brst\Psi$, where $\Psi$ is the respective string field. More precisely, if a
representation space is a space of functions with values in a graded space $\hat\cH$
with basis $e_A$ then the string field is the following object (see, {\it e.g.},~\cite{\BGST,Barnich:2003wj})
\begin{equation}
\Psi(x)=\psi^A(x) e_A \,, \qquad \gh{\psi^A}=-\gh{e_A}\,, \quad \p{\psi^A}=\p{e_A}\,,
\end{equation}
where $\psi^A$ are fields (including ghosts, antifields, etc) of the associated free field
theory determined by $s$. Note that $\gh{\Psi}=0$. The relation
$s\Psi=\hat\brst\Psi$ indeed defines the action of $s$ on fields $\psi^A$. This
action extends to space-time derivatives
$\d_{\mu_1}\ldots{\d_{\mu_k} }\psi^A$ through $\commut{s}{\dl{x^a}}=0$ and hence to
local functions (functions of fields and their derivatives).

It is useful to decompose the string field according to the ghost number of fields $\psi^A$
so that $\Psi=\dps\sum_k \Psi^{(k)}$ with $\Psi^{(k)}=\psi^{A_k}e_{A_k}\,,\,\,\gh{\psi^{A_k}}=k$.
The fields entering $\Psi^{(0)}$ are identified as physical fields. Gauge parameters are
associated with the fields entering $\Psi^{(1)}$.
The reducibility gauge parameters are then associated to fields entering $\Psi^{(2)}$ and
so on. The equations of motion and gauge symmetries are then
\begin{equation}
\hat\brst \Psi^{(0)}=0\,, \qquad \delta \Psi^{(0)}=\hat\brst \Psi^{(1)}\,, \qquad \delta \Psi^{(1)}=\hat\brst \Psi^{(2)}\,,\qquad \ldots\;,
%\qquad \hat\brst=\nabla+Q\,,
\end{equation}
where in the definition of the gauge transformations and the
reducibility relations one needs to replace ghost fields with the
respective gauge and reducibility parameters.

 It can be useful to
identify $\Psi^{(0)}$ with a general ghost-number-zero element of
the space of $\hat\cH$-valued functions. In the same way, the gauge parameters of order $l$ are identified with $\hat\cH$-valued functions of ghost number $-l+1$.\footnote{In the case where
physical fermionic fields are present this requires some care
because coefficients $\phi^A(x)$ in the expansion of a general
element $\phi=\phi^A e_A$ of the representation space are always
bosonic while the fields entering $\Psi=\psi^A e_A$
have Grassmann parity $\p{\psi^A}=\p{e_A}$.} For instance, in these
terms the gauge transformation law takes the usual form $\delta
\Psi^{(0)}=\hat\brst \xi^{(-1)}$, where $\xi^{(-1)}$ with
$\gh{\xi^{(-1)}}=-1$ is the gauge parameter.

Let us now explicitly find the field content, equations of motion and gauge symmetries of the theory determined by $\hat\brst$ and $\hat\cH$.
To this end, let us introduce the component fields entering the ghost-number-zero component of the string field
\begin{equation}
 \Psi^{(0)}=\psi_0+\psi_1+\ldots+\psi_{n-1}\,,\qquad
 \psi_p=\psi_{a_1\ldots a_{p}}^{i_1\ldots i_{p}}(x;y,a)b_{i_1}\ldots b_{i_{p}}\theta^{a_{1}}\ldots \theta^{a_{p}}\,.
\end{equation}
Fields $\psi_p$ are naturally identified as differential $p$-forms on $\manX$ taking values in the space of polynomials
in $y^a,a^a_i$ and ghosts $b_i$ subjected to the conditions \eqref{part-Weyl}. The equations of motion take the form
\begin{equation}
\begin{aligned}
 \nabla \psi_0+\sd_i\dl{b_i}\psi_1&=0\,,&\\
 \nabla \psi_1+\sd_i\dl{b_i}\psi_2&=0\,,&\\
&\ldots &\\%\ldots\,,\\
\nabla \psi_{n-1}&=0\,.&
\end{aligned}
\end{equation}
The gauge parameters corresponds to the ghost-number-one fields entering $\Psi^{(1)}$ and can be represented as
\begin{equation}
 \xi^{(-1)}=\xi_1+\xi_2+\ldots+\xi_{n-1}\,,\quad \xi_p=\xi^{i_1\ldots i_p}_{a_1\ldots a_{p-1}}(x;a,y)b_{i_1}\ldots b_{i_p}\theta^{a_1}\ldots \theta^{a_{p-1}}\;.
\end{equation}
For instance, gauge parameter  $\xi_1=\xi^ib_i$ and is a $0$-form. The gauge transformations have the form
\begin{equation}
\label{gs}
\begin{aligned}
 \delta_\xi \psi_0&= \sd_i\dl{b_i}\xi_1\,,&\\
 \delta_\xi \psi_1&= \nabla \xi_1+\sd_i\dl{b_i}\xi_2\,,&\\
\delta_\xi \psi_2&= \nabla \xi_2+\sd_i\dl{b_i}\xi_3\,,&\\
&\ldots &\\%\ldots\,,\\
 \delta_\xi  \psi_{n-1}&=\nabla \xi_{n-1}\,.&
\end{aligned}
\end{equation}
In the same fashion, one can also write down the reducibility parameters of order
$l$ that are associated to the fields  of ghost number $l+1$ and the respective reducibility
relations determined by $\hat\brst$. Note that in general there are fields of ghost number up
to $n-1$ so that there are reducibility relations of order up to $n-2$. The reducibility parameters
of order $l$ are $p$-forms with $p\leq n-3$.

The formulation determined by
\begin{equation}
\label{parent00}
\hat\brst\Psi^{(0)}=0\,, \quad \delta_\xi \Psi^{(0)}=\hat\brst \xi^{(-1)}\,, \quad \ldots\,, \quad \qquad \hat\brst=\nabla+Q\,
\end{equation}
is a natural generalization of the so-called intermediate form of the Fronsdal HS fields
found in~\cite{\BGST} (see also \cite{\BGadS} for the case of $AdS_d$ space) to the case of the mixed-symmetry fields.
Although this formulation appears here on the first place we keep the term ``intermediate'' because as we are going  to see
it is an intermediate formulation between the so-called parent formulation and the unfolded one.
In particular, for $n=2$ BRST operator $\hat\brst$ and hence the equations of motion and the gauge symmetries explicitly coincide with that identified in \cite{\BGST}.  We claim that \eqref{parent00} defines the gauge theory of mixed-symmetry HS fields on the Minkowski space.
Namely, we show that by eliminating the generalized auxiliary fields this theory can be taken to the explicitly Lagrangian form,
leading to Labastida equations of motion~\cite{Labastida:1986ft,Labastida:1989kw}.
In addition, eliminating a different collection of the generalized auxiliary fields one arrives at the unfolded form of the theory (this was recently constructed from scratch in \cite{Skvortsov:2008vs}).
Let us note that for the off-shell version of the Fronsdal theory the intermediate form naturally arises as a linearization of the nonlinear off-shell system in both Minkowski~\cite{Vasiliev:2005zu} and $AdS_d$ space \cite{Grigoriev:2006tt}.

Using the gauge symmetry \eqref{gs} one can always achieve $\bar
S^{\dagger i}\psi_0=0$, {\it i.e.} that $\psi_0$ takes values in
$\modM_0$ (cf. discussion before formula  \eqref{eqrel}). Moreover, in such a gauge
the first equation reduces to
$\nabla\psi_0=0$, where $\nabla$ is the Poincar\'e covariant
derivative acting in $\modM_0$. In this way one shows that in the
sector of $0$-forms the gauge system determined by $\hat\brst$
is indeed equivalent to \eqref{covconst-P}. In this sense, it can be understood as a gauge extension of the gauge
invariant formulation \eqref{covconst-P}. The difference is
similar (up to the auxiliary fields and extra gauge symmetries) to
the difference between Maxwell equations in terms of the curvature
$d*F=0, \,\,\, dF=0$ and the gauge description $d*F=0,\,\,F=dA$ in
terms of the potential.

As we will see in Section~\bref{sec:Q-coh} replacing \eqref{covconst-P} with \eqref{parent00} amounts, in particular, to replacing $\modM_0$ with the collection of the Poincar\'e modules $\modM_p\,,\,\,\,0\leq p\leq n-1$ appearing in the ghost-number-zero
$Q$-cohomology in the space \eqref{part-Weyl} tensored with the Grassmann algebra in new ghost variables $\theta^a$.
It is important to stress that already this step, in general, leads to the additional fields in the theory. Moreover,
more careful analysis shows that reducing to the $Q$-cohomology modifies the Poincar\'e covariant derivative entering equations of motion. Namely, it acts in the direct sum of modules in different degrees tensored with Grassmann algebra in
$\theta^a$ such that the Poincar\'e modules at different degrees are glued together. Note, that
the naive generalization of \eqref{covconst-P} (see Section \bref{sec:gen-brst} for more details)
would lead just to a collection of independent equations for fields at different degrees.
This phenomena is well-known in the unfolded description of the Fronsdal fields: the complete unfolded system contains
two type of fields -- HS connections and HS curvatures and the respective equations of motion are
not independent but related by the so-called central-on-mass-shell theorem
(see, {\it e.g.}, \cite{Vasiliev:2001wa}). As it was shown in~\cite{\BGST} this system can still be written in terms of just one module at the price of introducing a fiber BRST operator such that the two sets of fields appear in cohomology in different degrees while the central-on-mass-shell theorem is automatically built in. From this perspective
the present construction extends the one of \cite{\BGST} to the case of mixed-symmetry HS fields.

Let us also discuss the Poincar\'e invariance of the equations \eqref{parent00}. Because the Poincar\'e module for \eqref{parent00}
is just the space of all polynomials subjected to the Poincar\'e invariant conditions \eqref{part-Weyl} the Poincar\'e generators
have the usual form (without projectors, in contrast to \eqref{covconst-P}), {\it i.e.} the Poincar\'e symmetry acts on the fields
according to \eqref{hat-Poincare} with $P_a,M_{ab}$ represented on
polynomials in the standard way: $P_a=\dl{y^a},\,\, M_{ab}=y_a\dl{y^b}-y_b\dl{y^a}+a_{ai}\dl{a^b_i}-a_{bi}\dl{a^a_i}$. The
invariance is obvious because both $Q$ and the algebraic
conditions~\eqref{part-Weyl} are build from $sp(2n)$ generators
that by construction commute with the Lorentz generators $M_{ab}$.
They also commute with the Poincar\'e translations $P_a$.

It can be useful to define a different realization of Poincar\'e generators on the fields. Namely, the one
where $\theta^a$ and $x^a$
transform in the standard way
 \begin{equation}
 \label{bar-Poincare}
  \bar P_a \Phi=P_a\Phi\,,\quad \bar M_{ab}\Phi=M_{ab}\Phi+(x_a\dl{x^b}-x_b\dl{x^a})\Phi
   +(\theta_a\dl{\theta^b}-\theta_b\dl{\theta^a})\Phi\,.
 \end{equation}
Note that the transformation of $x^a$ implies that $dx^a$ also transform as Lorentz vectors, which
determines the transformation of $\theta^a$.

The realizations  (\ref{bar-Poincare}) and
(\ref{hat-Poincare}) differ by
$\hat\brst$-exact term
\begin{equation}
\bar M_{ab} -\hat M_{ab}=\commut{{\hat\brst}}{x_a \dl{\theta^b}-x_b \dl{\theta^a}}\,.
\end{equation}
This, in particular, implies that $\commut{\hat\brst}{\bar
M_{ab}}=0$, because that is true for $\hat M_{ab}$. Moreover, it also implies that these two representations are equivalent.
Indeed, global symmetries of the theory are ghost-number-zero
operators commuting with $\hat\brst$ while those in the image of
the adjoint action are trivial symmetries (on-shell equivalent to
the gauge symmetries). It follows that inequivalent symmetries are
operator $\hat\brst$-cohomology at zeroth ghost degree. In
particular, $\hat M_{ab}$ and $\bar M_{ab}$ are different
representatives of the same cohomology class. Note
that one can also take as $\bar P_a=\dl{x_a}$ by adding $\commut{\hat\brst}{\dl{\theta^a}}$.

\subsection{Standard Lagrangian BRST first-quantized formulation}
\label{sec:standard}

Given a theory of the form  \eqref{parent00} determined by $\hat \brst$ one can easily eliminate
variables $y^a$ and $\theta^a$ in order to end up with the standard first-quantized BRST
description (see \cite{\BGST} for more details). However, this is only possible if no constraints
involving $\dl{y^a}$ are imposed on $\Psi$ because such constraints become differential in $x^a$
once $y^a$ are eliminated. At the case at hand $\Psi$ takes values in the space of elements
annihilated, in particular, by
\begin{equation}
\label{additconstr}
\Box_y\equiv T^{00}=\dl{y_a}\dl{y^a}\,,\qquad
 S^i\equiv T^{i0}=\dl{a_{ia}}\dl{y^a}\,.
\end{equation}
These are constraints from \eqref{part-Weyl} that involve $\dl{y^a}$.
The way out is to impose these constraints through the BRST procedure.

To this end,
one introduces additional Grassmann odd ghost variables $c_0,\bar b^0,c_i,\bar b^i$
satisfying
\begin{equation}
 \commut{\bar b^0}{c_0}=1\,,\quad \commut{\bar b^i}{c_j}=\delta^i_j\,,\qquad  \gh{c_i}=\gh{c_0}=1\,,\quad
 \gh{\bar b^i}=\gh{\bar b^0}=-1\;,
\end{equation}
and represented on polynomials in $c_0,c_i$ so that $\bar b^i\phi=\dl{c^i}\phi$ and $\bar b^0\phi=\dl{c^0}\phi$.
The extended BRST operator is given by
\begin{equation}\label{parent1}
\brst^{\rm parent}=\theta^a(\dl{x^a}-\dl{y^a})+c_0\Box_y +c_i S^i+S^{\dagger}_ i \dl{b_i}-c_i\dl{b_i}\dl{c_0}\;.
\end{equation}
As a representation space one takes $\cH^{\rm parent}$-valued functions in $x^a$, where
$\cH^{\rm parent}$ is a tensor product of Grassmann algebra in $\theta^a$ with polynomials
in $y,a_i$ and ghosts $c_0,c_i,b_i$ subjected to the appropriate modification of the  remaining constraints $T^{ij}\;,\; \hat\cN_i{}^j\,\,\; i>j\;,\; \hat\cN_i-s_i$.
More precisely, these constraints are modified by the ghost contributions needed to maintain
their BRST invariance with respect to the extended BRST operator~\eqref{parent1}
and are given explicitly by
\be
\begin{gathered}
\label{const-ext}
\cT^{ij}=\eta^{mn}\dl{a^m_i}\dl{a^n_j}+\dl{c_i}\dl{b_j} +\dl{c_j}\dl{b_i}\,,\qquad
\cN_i{}^j=a^a_i\dl{a^a_j}+c_i\dl{c_j}+b_i\dl{b_j}\;,\\
\cN_i=a^a_i\dl{a^a_i}+c_i\dl{c_i}+b_i\dl{b_i}\quad \text{(no summation over $i$)}\;.
\end{gathered}
\ee
Note that they indeed do not involve $\dl{y^a}$. Let us also note that these constraints are BRST extensions of the generators of the upper triangular subalgebra and the Cartan elements of $sp(2n-2)\subset sp(2n)$ algebra that is a Howe dual to the Lorentz algebra acting on the space of variables $a^a_i$.
\begin{prop}
Parent system $(\brst^{\rm parent}, \cH^{\rm parent})$ and
intermediate system $(\hat\brst,\hat\cH)$ are equivalent.
\end{prop}
\begin{proof}
To prove the proposition one introduces a grading defined by a homogeneity degree in $c_0$ and $c_i$.
Then using the method of the homological reduction described in Appendix~\bref{sec:C} one finds that
$\brst^{\rm parent} = \brst^{\rm parent}_{-1}+ \brst^{\rm parent}_0$, where $\brst^{\rm parent}_{-1}=c_0\Box_y +c_i S^i$ and
the theory can be reduced to the cohomology of $\brst^{\rm parent}_{-1}$.
We now need to invoke the homological result demonstrated in Appendix~\bref{sec:A}.
Namely, the crucial fact is the cohomology of the operator $\Delta_{IJ}=C_{IJ}T^{IJ}$ (no summation
over $I,J$),
where $C_{IJ}$ are some ghost variables, in the space of all polynomials is given by
$C_{IJ}$-independent elements annihilated by $T^{IJ}$. In particular, the cohomology of
$\brst^{\rm parent}_{-1} = C_{00}T^{00}+C_{0i} T^{0i}$ is concentrated in the zeroth
degree and can be identified with $c_0, c_i$-independent elements annihilated
by $T^{00}$ and $T^{0i}$. It follows that the cohomology subspace is singled out
by constraints \eqref{part-Weyl}, because for $c_0, c_i$-independent elements
constraints \eqref{const-ext} along with  $T^{00}$ and
$T^{0i}$ explicitly coincide with  \eqref{part-Weyl}.
The reduced BRST operator
coincides then with $\brst^{\rm parent}_0$ restricted to the cohomology, and
hence coincides with $\hat\brst$.
\end{proof}

Because the constraints \eqref{const-ext} determining the representation space for the parent
system are algebraic (do not involve $\dl{y^a}$) the
elimination of $y^a,\theta^a$ variables is now straightforward and amounts to dropping the
first term (the one with $\theta^a$) and replacing $\dl{y^a}\to \dl{x^a}$ in the remaining terms.
This can be seen by reducing the theory to the cohomology of $\theta^a\dl{y^a}$ and evaluating
the reduced BRST operator (see \cite{\BGST} for more details and the proof).
In this way one arrives at the theory determined by the following
BRST operator
\begin{equation}
\label{standard-brst}
\brst=c_0\Box +c_i \cS^i+\cS^{\dagger}_ i \dl{b_i}-c_i\dl{b_i}\dl{c_0}\,,
\end{equation}
where the constraints are given by
\begin{equation}
 \Box=\eta^{ab}\dl{x^a}\dl{x^b}\,,
 \qquad \cS^\dagger_i=a^a_i \dl{x^a}\,,
 \qquad \cS^i=\dl{a^a_i}\dl{x_a}\,,
\end{equation}
and satisfy the following algebra:
\be
\label{const1line}
\ba{c}
[\Box, \cS_i]  =[\Box, \cS^\dagger_i] = 0\;, \quad [\cS_i, \cS^\dagger_j] =\delta_{ij}\,\Box\;,
\qquad
[\cS_i, \cS_j] = [\cS^\dagger_i, \cS^\dagger_j]= 0\;.
\ea
\ee

The representation space is now given by $\cH$-valued functions in $x^a$, where $\cH$
is the space of polynomials in $a_i$ and ghosts $c_0,c_i,b_i$ satisfying
\begin{equation}
\label{brstextconst}
\cT^{ij}\phi=0\,,\qquad
\cN_i{}^j\phi=0\;\;\;i>j\,,\qquad
\cN_i\phi=s_i\,\phi\,,
\end{equation}
where the BRST invariant constraints $\cT^{ij}, \cN_i{}^j$ and $\cN_i$ are defined in
\eqref{const-ext}. In this way we have arrived at the following
\begin{prop}
System $(\brst,\cH)$ is equivalent to
$(\brst^{\rm parent}, \cH^{\rm parent})$ and hence to intermediate system $(\hat\brst,\hat\cH)$.
\end{prop}

As we are going to see system $(\brst,\cH)$
yields  a Lagrangian description of mixed-symmetry fields on the Minkowski space.
In particular, we explicitly show that the appropriate reduction of this theory gives the theory proposed by Labastida in \cite{Labastida:1986ft,Labastida:1989kw}.
In addition, we also show in Section \bref{sec: unfolding}
that the parent theory \eqref{parent1} (as well as the equivalent formulation \eqref{parent00}) contains the unfolded
form of this model proposed in \cite{Skvortsov:2008vs}.
As an independent check, in Section~\bref{sec: Wigner} we observe that
system $(\brst,\cH)$ indeed describes the irreducible massless unitary
representation of the Wigner little group~\cite{Wigner:1939cj}
determined by the spins $s_{n-1},\,\ldots,\, s_1$.

There is an alternative motivation for implementing the constraints $S^i$ and $\Box$ \eqref{additconstr}
through the BRST operator.
Namely, it turns out that the BRST operator \eqref{standard-brst} is
symmetric with respect to the standard inner product
of the form
\begin{equation}
\label{innerprod}
\inner{\phi}{\psi}=\int d^dx \int dc_0\;\inner{\phi}{\psi}_0\,,
\end{equation}
where $\inner{}{}_0$ is the standard Fock inner product in the space of polynomials in $c_i,b^i,a^a_i$ identified with the
Fock space generated by $c_i,b^i,a^a_i$ from the vacuum state $\vac$ defined by
$\bar a \vac=\bar c\vac=\bar b\vac=0$.
In our notations the conjugation rules have the following form
\begin{equation}
\label{conj}
(a_i^a)^\dagger=\bar a_i^a\,,\qquad (b_i)^\dagger=-\bar b^i \,,
\qquad (c_i)^\dagger=\bar c^i\,,
\end{equation}
where the space-time and the internal indices are raised and lowered with
the Minkowski metric $\eta_{ab}$ and the standard
Euclidean metric $\delta_{ij}$ on the internal space, respectively.
This can be seen as equipping the Grassmann odd superspace of ghosts with the super-Euclidean
metric $\epsilon_{\alpha\beta}\delta^{ij}$. Indeed, introducing the collective notation $\chi^\alpha_i$
for $\chi_i^1=c_i$ and $\chi_i^2=b_i$ one finds $\inner{\chi^\alpha_i}{\chi^\beta_j}_0=\epsilon^{\alpha\beta}\delta_{ij}$
(here, $\epsilon^{\alpha\beta}$ is a $2d$ Levi-Civita symbol defined such that
$\epsilon_{12}=\epsilon^{12}=1$).
One can check that this is indeed consistent with the commutation relation~\footnote{We use
 the following convention for the conjugation in the presence of fermions:
$(ab)^\dagger=(-1)^{\p{a}\p{b}}b^\dagger a^\dagger$ and $\inner{\phi}{a \psi}=(-1)^{\p{a}\p{\phi}}\inner{a^\dagger \phi}{\psi}$, where $\p{a}$ denotes the Grassmann parity of $a$.}
and uniquely determines the inner product for which
$\dagger$ is the hermitian conjugation. The inner product just defined carries ghost degree $-1$, {\it i.e.} $\inner{\phi}{\psi}=0$ for any $\phi,\psi$ such that $\gh{\phi}+\gh{\psi}\neq 1$.

Given a nondegenerate inner product of ghost number $-1$ and a symmetric nilpotent BRST operator one immediately constructs
the action\footnote{The choice of the overall sign corresponds to ``almost-positive'' signature of $\eta_{ab}$.}
\begin{equation}
\label{action}
 S=\half\inner{\Psi^{(0)}}{\brst \Psi^{(0)}}
\end{equation}
that determines the equations of motion $\brst \Psi^{(0)}=0$.  The Batalin--Vilkovisky master action
of the theory can be constructed  in the form $S^{BV}=\half\inner{\Psi}{\brst \Psi}$ so that the fields entering the
nonzero components of the string field $\Psi$ are naturally identified with antifields and ghost fields of the BV formalism.

\bigskip

To make a contact to the literature and to highlight the quantum
mechanical interpretation of the system
let us sketch the formulation of the constrained system whose BRST operator is given by $\brst$. The variables are as follows.
The space-time variables $x^m$, $p_n$ satisfy canonical commutation
relations and conjugation rules
\begin{equation}
[p_m, x^n] = -\delta_m{}^n\,,\qquad (x^m)^\dagger=x^m\,,\quad p_n^\dagger=-p_n\,.
\end{equation}
Internal (spin) variables (oscillators) satisfy
\begin{equation}
[\bar a_i^m,a_j^{n}] = \delta_{ij}\eta^{mn}\,,\qquad (a_i^m)^\dagger=\bar a_i^m
\qquad i,j = 1,2,\ldots,n-1\,.
\end{equation}

The constraints are:
\footnote{Related constraints were discussed in~\cite{Hallowell:2007qk}
in the context of constant curvature spaces.}
\begin{gather}
\Box\,,\qquad \cS^\dagger_i\,,\qquad \cS^i,\\
T^{ij}\,,\qquad N_i{}^j\;\;\,\,\,i>j\,, \qquad N_i-s_i\,.\label{2nd-line-const}
\end{gather}
All together these constraints form an upper-triangular subalgebra of $sp(2n)$
along with $n-1$ Cartan elements (weight conditions for spin oscillators).
The second line form the subalgebra of $sp(2n-2)\subset sp(2n)$
that do not affect the space-time variables $x^a$.~\footnote{In the case of $n=3$, \textit{i.e.}, for two-row covariant fields,
this $sp(4)$ algebra has been identified in \cite{Burdik:2001hj}.} The first line contains those constraints that do involve space
time derivatives. These constraints also form a subalgebra \eqref{const1line}.

The first-class constraint quantum system is defined by
implementing the first line through the BRST operator using the
ghost variables $\bar b^0,c_0$, $\bar b^i, c_i$ and $\bar c^i,
b_i$ while imposing the BRST invariant extensions of the
constraints from the second line directly in the representation
space. The BRST operator for this system is indeed $\brst$ given
by~\eqref{standard-brst}.
 The respective representation $\cH$ space is singled out by
the conditions \eqref{brstextconst} that are BRST invariant extensions of the
constraints \eqref{2nd-line-const}.

It is important to stress that one can as well consider the theory
determined by the action \eqref{action} where $\Psi^{(0)}$ is not required
to satisfy constraints \eqref{2nd-line-const}. Such theory describes
a direct sum of irreducible fields, where any field enters the theory with
(in general infinite) multiplicity. With the constraints
\eqref{2nd-line-const} relaxed, action \eqref{action} is known in the literature
and can be identified as an appropriate truncation of the
open bosonic string field theory Lagrangian in the tensionless
limit~\cite{Sagnotti:2003qa,Bonelli:2003kh}.

\section{Relation to the Labastida approach}\label{sec:Labastida}

In this section we analyze the dynamical content of the theory determined by
the BRST operator $\brst$ \eqref{standard-brst}. We explicitly prove that the equations
of motion for mixed-symmetry fields generated by the BRST operator $\brst$
coincide with those originally obtained by Labastida \cite{Labastida:1986ft, Labastida:1989kw}.

\noindent Let us sketch  the main features of the Labastida equations
formulated for an individual spin field with $n-1$ rows.
The kinetic operator $L$ has the form
\begin{equation}
\label{laboper}
L = \Box - D_i D^i + \frac{1}{2}\,D_i D_j \;T^{ij} \;,
\end{equation}
where $D_i = a^a_i\dl{x^a}$ and $D^i = \bar a^{i}_a\dl{x^a}$, $i=1,...,n-1$ and the trace annihilation $T^{ij}$ operator
is defined by (\ref{SPgenerators}).
The operator $L$ obviously commutes with Young symmetrizers $N_i{}^j$ (\ref{newnot}), which
implies that the Young symmetry properties of the fields are shared by the
operator $L$. Fields $\varphi$ satisfy the  following trace constraint
\be
\label{Labtr}
T^{(ij}T^{kl)}\varphi = 0\;
\ee
that singles out double-traceless fields. The Labastida equations of motion $L\varphi=0$
are invariant under the gauge transformations
\be
\label{gtrLab}
\delta \varphi = D_i \Lambda^i\;,
\ee
provided that parameters satisfy
\be
\label{trconLab}
T^{(ij}\Lambda^{k)} = 0\;.
\ee
Let us stress that in general a parameter $\Lambda^i$ for a given index $i$ does not satisfy Young symmetry conditions.
Instead, parameters for different $i$ and $j$ are related to each other by the appropriate Young symmetrizations.
Such relations can be easily read off from (\ref{gtrLab}) by imposing Young symmetry condition on the left-hand-side.
It turns out that representing the space of gauge parameters by tensors $\Lambda^i$ not satisfying
Young symmetry conditions allows one to write down the gauge transformation law in a simple form
\footnote{This property of the Labastida approach was originally observed in \cite{Burdik:2001hj}
within the BRST formulation of two-row field dynamics in Minkowski space.}.

\subsection{Polynomials in ghosts and associated algebras}

Before considering a dynamics described by %the  string field and the
BRST operator $\brst$ let us discuss the Fock subspace generated
by ghost variables $c_i,b_i$ from the more algebraic point of
view. Introducing a collective notation  $\chi_i{}^\alpha = (c_i,
b_i)$, $\alpha=1,2$ for ghost variables it is convenient to
consider $\chi^\alpha_i$ as coordinates on the tensor product of
two superspaces with bases $e^i$ and $e_\alpha$. This tensor
product is equipped with the metric
$\epsilon_{\alpha\beta}\delta^{ij}$
that makes a Euclidian
superspace. Recall that it is this metric that induces the inner
product on the Fock space of ghost variables (see Section
\bref{sec:standard}). Note that this metric factorizes into the
super-Euclidian metric $\epsilon_{\alpha\beta}$ and the
supersymplectic metric $\delta^{ij}$. Using the
$\epsilon_{\alpha\beta}$ factor allows one to introduce oscillator
realizations of $sp(2)$ and $sp(2n-2)$ algebras. These algebras
provide convenient tools for the analysis of the dynamical content
of the theory.

Remarkably, ghost variables introduce into the game one more Howe dual pair which is complementary to the previous  one considered in Section \bref{sec: howe}. These new dual algebras
are $gl(n-1)$ and $gl(2)$ and their generators are given by
\footnote{This construction also enjoys a supersymmetric extension. To this
end we note that $\chi_i{}^\alpha$ transforms both as $gl(n-1)$ and $gl(2)$ vectors
and hence we can build supercharges
$$
Q_a^\alpha = a_a^i \chi^\alpha_i\;,
\quad
Q_\alpha^a = \dl{a_a^i} \dl{\chi^\alpha_i}\;,
\qquad
\{Q_a^\alpha, Q_\beta^b\} = \delta^\alpha_\beta L_a{}^b-\delta_a^b Z^\alpha{}_\beta\;,
$$
where $ L_a{}^b = \half \, \{a_a{}^i, \dl{a_b{}^i}\}$ are $gl(d)$ generators.
The resulting superalgebra is $gl(d|2)$.
}
\be
Y_i{}^j = \half\,\Big(\chi_i{}^\alpha \dl{\chi_j{}^\alpha}-\dl{\chi_j{}^\alpha}\chi_i{}^\alpha \Big)\;,
\qquad
Z^\alpha{}_\beta = \half\,\Big(\chi_i{}^\alpha \dl{\chi_i{}^\beta}-\dl{\chi_i{}^\beta}\chi_i{}^\alpha \Big)\;.
\ee

Using the $\epsilon^{\alpha\beta}$ factor the algebra $gl(n-1)$ can be extended by the following generators
\be
Y_{ij}=\epsilon_{\alpha\beta}\chi_i{}^\alpha \chi_j{}^\beta\;,
\qquad
Y^{ij} = \epsilon^{\alpha\beta} \dl{\chi_i{}^\alpha} \dl{\chi_j{}^\beta}\,,
\ee
so that similarly to \eqref{SPgenerators} generators $Y_{ij},Y_i{}^j$, and $Y^{ij}$ form $sp(2n-2)$ algebra.
In particular, BRST extended algebraic operators  (\ref{const-ext}) can be represented as
\be
\cT^{ij} = T^{ij}+Y^{ij}\;,
\qquad
\cN_i{}^j = N_i{}^j+Y_i{}^j+\delta_i{}^j\;.
\ee
Analogously one introduces the algebra $sp(2)\simeq sl(2)$ generated by
\be
Z_{\alpha\beta}  = \epsilon_{\alpha\gamma}Z^\gamma{}_\beta+\epsilon_{\beta\gamma}Z^\gamma{}_\alpha\;.
\ee
The standard basis of $sp(2)$ algebra  reads
\be
\label{fermsl}
\ba{c}
\dps
Z_+ \equiv Z^1{}_2 = c_i \dl{b_i}\,,
\quad
Z_- \equiv Z^2{}_1= b_i \dl{c_i}\;,
\\
\\
\dps
Z_0 \equiv Z^1{}_1 - Z^2{}_2 =c_i \dl{c_i}- b_i \dl{b_i}\;.
\ea
\ee
%In particular, it shows the isomorphism between $sp(2)$ and $sl(2)$ algebras.

\subsection{The ghost-number-zero fields}

It is convenient to represent string field $\Psi\equiv\Psi(a,b,c |x)$
as follows
\be
\label{triplet}
\Psi= \Psi_1 +c_0 \Psi_2\;,
\ee
For the ghost-number-zero component $\Psi^{(0)}$ fields
$\Psi_1^{(0)}\equiv \Phi$ and $\Psi_2^{(0)}\equiv C$ are the following decompositions with respect to the ghost variables:
\be
\label{sf}
\ba{l}
\dps
\Phi =
\sum_{k=0}^{n-1}\;c_{i_1} \cdots c_{i_k} b_{j_1} \cdots b_{j_k}\,
\Phi^{i_1...i_k|j_1...j_k}\;,
\\
\\
\dps
C=\sum_{k=0}^{n-2}\;c_{i_1} \cdots c_{i_k}
b_{j_1} \cdots b_{j_{k+1}}\,
C^{i_1...i_k|j_1...j_{k+1}}\;.

\\
\ea
\ee
The expansion coefficients  $\Phi^{i_1...i_k|j_1...j_k}(a|x)$
and $C^{i_1...i_k|j_1...j_{k+1}}(a|x)$ are $gl(n-1)$ tensors antisymmetric in each group
of indices, and the slash $|$ implies that no symmetry properties between two groups of indices
are assumed. In the sequel we use the notation $\varphi$ for the $k=0$ component of $\Phi$.
Note that component fields $\Phi^{i_1...i_k|j_1...j_k}$
and $C^{i_1...i_k|j_1...j_{k+1}}$ were considered in
\cite{Sagnotti:2003qa}. These can be seen
as a generalization of the so-called triplet originally discussed in
\cite{Bengtsson:1986ys} in the context of totally
symmetric fields.

Our aim now is to find a minimal set of fields that covariantly describes an individual
spin field. This is achieved in two steps. As a first step we eliminate all the generalized auxiliary fields
entering the formulation determined by $\brst$. This is achieved using the general method of~\cite{Barnich:2004cr}.
As a second step we subject the string field $\Psi^{(0)}$ to the remaining irreducibility conditions, namely, the
Young symmetrizer, the trace conditions, and the weight conditions.

\subsection{$\brst_{-1}$ cohomology}
\label{Sec:Cohomology}

Let us decompose  the BRST operator with respect to the homogeneity degree in $c_0$ as
\begin{equation}
\label{decbrstLab}
\brst=\brst_{-1}+\brst_{0}+\brst_{1}\,,
\end{equation}
with
\begin{equation}
\label{3brst}
  \brst_{-1}=-c_i\dl{b_i}\dl{c_0}\,,\quad
\brst_0=c_i \cS^i+\cS^{\dagger}_ i \dl{b_i}\,,\quad
\brst_1=c_0\Box\;.
\end{equation}
The lowest degree component $\brst_{-1}$ is purely algebraic so that all the fields that are not in the
cohomology of $\brst_{-1}$
are generalized auxiliary fields (see \cite{\BGST} for details).

In order to analyze the cohomology of $\brst_{-1}$ let us note that it can be represented
in the form
\be
\label{-1rep}
\brst_{-1} =  -Z_+ \,\dl{c_0}\;,
\ee
where $Z_+$ is a generator of  $sl(2)$ algebra realized on ghost fields (\ref{fermsl}).

We are now going to find $\brst_{-1}$ cohomology in the subspace
of elements satisfying $\cT^{ij}\phi=0\,,\,\,\; \cN_i{}^j\phi=0\,
\,\,\, i>j$ and $\cN_i\phi=(s_i+\frac{d}{2})\phi$. To this end it is
useful to identify first the cohomology in the entire
representation space and then impose the conditions. This is
legitimate because of the following argument: the conditions we
are dealing with are the highest weight conditions for the
$sp(2n-2)$ algebra formed by $\cT^{ij},\cN^i{}^j, \cT_{ij}$, where $\cT_{ij}$
is a BRST invariant extension of $T_{ij}$.
Decomposing the entire space into the direct sum of irreducible
highest weight $sp(2n-2)$-modules one finds that any element can
be represented as a sum of elements of the form
$\phi=\phi_0+\cT_A\phi^A$, where $\phi_0$ satisfy the highest
weight conditions and $\cT_A$ is a collective notation for all the
generators from the lower-triangular subalgebra ({\it i.e.} $
\cT_{ij},\cN_i{}^j\,\;\; j>i$). Because $sp(2n-2)$ commutes with
$\brst_{-1}$ one concludes that $\brst_{-1}$ does not  map elements
of the form $\cT_A\phi^A$ to elements satisfying highest
weight conditions. This implies that the coboundary condition is
not affected by restricting to the subspace so that
$\brst_{-1}$  cohomology in the subspace coincides with the
restriction to the subspace of the $\brst_{-1}$ cohomology in the
entire space.

Using the representation \eqref{-1rep} the searched-for cohomology
$H(\brst_{-1})$ in the entire space can be readily found (see also Section \bref{Sec: sl(2)}).
Indeed $\im{\brst_{-1}}$ is given by
$c_0$-independent elements that are in the image of $Z_+$. It follows that one can represent the $c_0$-independent
cohomology by elements annihilated by $Z_-$. Let us consider then  $\Ker{\brst_{-1}}$ for $c_0$-dependent elements (for $c_0$-independent the cocycle condition is satisfied trivially).
It follows that elements annihilated by $Z_+$ satisfy the cocycle condition and are in cohomology.
Decomposing a general element $\phi$ into $c_0$-(in)dependent elements according to $\phi = \phi_1 + c_0 \phi_2$ we can formulate
cohomological conditions as $Z_- \phi_1=0$ and $Z_+ \phi_2 = 0$. Eigenvalues of the generator $Z_0$
are integer numbers, $Z_0 \phi_1 = m \phi_1$ and $Z_0 \phi_2 = (m-1)\phi_2$,
where $m=gh(\phi)$ is a ghost number.
For instance, for $m=0$ we obtain $\phi_2=0$ and $Z_{\pm}\phi_1=0$ and, hence, $c_0$-independent component is $sl(2)$ invariant.
Let us also note that the choice of representatives is consistent with the conditions
(\ref{brstextconst}) because both $Z_+$ and $Z_-$ commute with (\ref{brstextconst}).
This determines the structure of the physical fields entering $\Psi^{(0)}$.

For positive values of the ghost number $+m$, $0\leq m\leq n-1$ the cohomology  $H^{m}$
is given by elements $\phi_1 = 0$ and $Z_+\phi_2=0$. This implies that the string field $\Psi^{(-m)}$ takes the following form
\be
\label{cohC}
\Psi^{(-m)}=\Psi_2^{(-m)} =\sum_{k=0}^{n-m-2}\;c_{i_1} \cdots c_{i_{k+m}}
b_{j_1} \cdots b_{j_{k+1}}\,
\Psi_2^{i_1...i_{k+m},\,j_1...j_{k+1}}\;,
\ee
where all components $\Psi_2^{i_1...i_{k+m},\,j_1...j_{k+1}}$ are $gl(n-1)$
Young tableaux with columns
of heights $k+m$ and $k+1$.
Elements $\Psi_2^{(m)}$ of the cohomology are irrelevant in the present analysis and correspond to the antifields
of the Batalin-Vilkovisky formulation of the theory.

For non-positive values of the ghost number $-m$, $0\leq m\leq n-1$ the cohomology $H^{-m}$ is
given by elements with $\phi_2=0$ and $Z_-\phi_1=0$, so that the string field takes the form
\be
\label{fincoh}
\Psi_1^{(m)} =\sum_{k=0}^{n-m-1} \,
c_{i_1} \cdots c_{i_k} b_{j_1} \cdots b_{j_{k+m}}
\Psi_1^{j_1\ldots j_{k+m},\,i_1\ldots i_k}\;,
\ee
where all components $\Psi_1^{j_1\ldots j_{k+m},\,i_1\ldots i_k}$ are $gl(n-1)$ Young tableaux with columns
of heights $k+m$ and $k$.
The expansion coefficients of $\Psi_1^{(m)}$ are identified with dynamical fields (at $m=0$)
and ghost field of $(m-1)$-th level of reducibility (for $m\neq 0$) associated to the respective gauge parameters.
Recall that in addition one needs to impose the conditions (\ref{brstextconst}) in order to describe
cohomology in the subspace.

More detailed discussion of the gauge symmetries of the theory will be given in Section \bref{Sec: GT and EOM}.

\subsection{BRST extended algebraic conditions}

Let us now analyze algebraic irreducibility conditions (\ref{brstextconst}) imposed on the
representatives of $\brst_{-1}$ cohomology.
Representing the string field as $\Psi = \Psi_1+c_0 \Psi_2$ the BRST extended trace constraint (\ref{const-ext})
takes the form
\be
\label{exttrcon}
(T^{ml}+Y^{ml})(\Psi_1+c_0 \Psi_2)=0\;.
\ee
Obviously, it does not mix up traces of $\Psi_1$ and $\Psi_2$ and hence they can be analyzed separately.
In both sectors constraint (\ref{exttrcon}) relates $k+1$-th component of the cohomology
to the trace of $k$-th one. Applying $T^{ps}$ and $Y^{ps}$ to the left-hand-side of the above expression yields the relation $T^{ml}T^{ps}\Psi_{1,2} = Y^{ml}Y^{ps}\Psi_{1,2}$. Observing  then  that a
symmetrized combination $Y^{(ml}Y^{ps)}$ is identically zero one obtains that $\Phi$ satisfies the double
trace constraint $T^{(ml}T^{ps)}\Psi_{1,2} = 0$.

For the dynamical fields associated to the lowest component of the cohomology $H^{0}$  we recover the
familiar Labastida constraint (\ref{Labtr})
\be
\label{dtr}
T^{(ps}T^{ml)}\varphi = 0
\ee
and find that all other components $\Phi^{j_1\ldots j_{k},\,i_1\ldots i_k}$ for $k>0$ are expressed in terms of
the traces of $\varphi$.
For instance, for lowest values of $k$ the corresponding expressions read off from  $T^{ml}\Phi + Y^{ml}\Phi=0$
are
\be
\label{trrel1}
T^{ml}\varphi-\Big(\Phi^{l|m}+\Phi^{m|l}\Big) = 0\;,
\ee
\be
T^{ps}T^{ml}\varphi +4\Big(\Phi^{sl|pm}+\Phi^{pl|sm}+\Phi^{sm|pl}+\Phi^{pm|sl} \Big) = 0\;.
\ee
Identifying the right-hand-sides with appropriate symmetrizations of   $\Phi^{m,l}$ and
$\Phi^{sl,\,pm}$ respectively,  we obtain
formulas that express these components through the field $\varphi$.

The cohomology $H^{-1}$ that corresponds to gauge parameters of the zeroth  level
can be analyzed along the same lines. In particular, for $k=0$ component $\Lambda^i\equiv \Psi_1^i$
of $H^{-1}$ one obtains the relation
\be
T^{(mn} \Lambda^{i)} =0\;
\ee
which is Labastida constraint (\ref{trconLab}) for the gauge parameters. Quite analogously to the
dynamical fields, higher order components of $H^{-1}$ are expressed via traces of the gauge parameter
$\Lambda^{i}$.

To analyze Young symmetry types of the fields we impose the BRST extended algebraic  conditions
\be
\label{condit1}
(N_i{}^j+Y_i{}^j)(\Psi_1+c_0 \Psi_2) = 0 \quad i>j\;,
\ee
and
\be
\label{condit2}
(N_i + Y_i-s_i-1)(\Psi_1+c_0 \Psi_2) = 0\;,
\ee
where $s_i$ are  integer spins and $Y_i\equiv Y_i{}^i$ for a fixed $i$. In particular,
for the field $\varphi$ we obtain $N_i{}^j \varphi = 0 \quad i>j$, {\it i.e.}, it is
described by Young tableau of the type $(s_{n-1}, s_{n-2}, \cdots, s_1)$ and the corresponding
gauge parameter $\Lambda^i$ has one less $i$-th oscillator $a_i^a$. Let us note that
BRST extended conditions (\ref{condit2}) do not in general lead to Young symmetries of $\Lambda^i$.
Instead there appears a set of recurrent relations between $\Lambda^i$ generated by Young symmetrizers
$N_i{}^j\;\;\;i>j$. Their form can be easily read off from (\ref{condit1}).

Finally, let us note that the representative of the cohomology $H^{-(n-1)}$
has the following form $\Psi_1^{(n-1)}=b_{1} \cdots b_{{n-1}} \Psi_1^{1\ldots {n-1}}$, {\it i.e.}
corresponds to the maximally antisymmetric tensor.
As discussed above it corresponds to the gauge parameter of the maximal depth of reducibility  $n-1$. The conditions
\eqref{condit1}, \eqref{condit2} applied to $\Psi_1^{(n-1)}$ reduce to $N_i{}^j \Psi_1^{(n-1)}=0\;\;\;i>j$
and $N_i\Psi_1^{(n-1)} = (s_i-1)\Psi_1^{(n-1)}$. In terms of Lorentz irreps it corresponds
to Young tableau with one leftmost column cut off compared to the tableau associated with the dynamical
field $\varphi$.

\subsection{Gauge transformations and field equations}
\label{Sec: GT and EOM}

In order to describe the theory reduced to $\brst_{-1}$-cohomology $H$ one is to
compute the reduced operator $\tilde\brst$ acting in $H$. $\tilde\brst$  determines the equations of motion, gauge symmetries, and reducibility relations of the reduced theory and can be found using the standard cohomological technique
(see \cite{\BGST} for an exposition in the similar terms). We now take a different route and obtain the explicit form of the reduced equation of motion and gauge symmetries by explicitly eliminating the generalized auxiliary fields associated to the contractible pairs
for $\brst_{-1}$.

The gauge transformations
\be
\label{gaugebrst}
\delta  \Psi^{(0)}   = \brst\, \xi^{(-1)} \;,
\qquad
\gh{\xi^{(-1)}}=-1\;.
\ee
involve the gauge parameters of the form $\xi^{(-1)} = \Lambda + c_0 \Upsilon$, where
$\gh \Lambda = -1$ and $\gh \Upsilon=-2$, and
\be
\ba{l}
\dps
\Lambda = \sum_{k=0}^{n-2}\;c_{i_1} \cdots c_{i_k} b_{j_1} \cdots b_{j_{k+1}}\,
\Lambda^{i_1...i_k|j_1...j_{k+1}}\;,
\\
\\
\dps
\Upsilon = \sum_{k=0}^{n-3}\;c_{i_1} \cdots c_{i_k} b_{j_1} \cdots b_{j_{k+2}}\,
\Upsilon^{i_1...i_k|j_1...j_{k+2}} \;.
\ea
\ee
The gauge symmetry is reducible and there exists the set of level-$(l-1)$ ($1\leq l \leq n-1$) gauge parameters
and gauge transformations of the form
\be
\delta  \xi^{(-l)}   = \brst\, \xi^{(-l-1)}\;,
\qquad \gh{ \xi^{(-l)}} = -l\;.
\ee
Recall that gauge parameters are also subjected to the BRST extended irreducibility conditions~\eqref{brstextconst}.

For fields $\Phi$ and $C$ the gauge transformations take the form
\be
\ba{l}
\delta \Phi  = Z_+ \Upsilon +\brst_{0}\Lambda\;,
\\
\\
\delta C = \Box \Lambda - \brst_0 \Upsilon\;.
\ea
\ee

We observe that the transformation for fields $\Phi$ contains an algebraic term $Z_+\Upsilon$.
It means precisely that the part of components of fields $\Phi$ are Stueckelberg-like and can be gauged  away
by imposing the proper gauge condition. Using the cohomological analysis of Section \bref{Sec:Cohomology}
we conclude that the remaining components of fields $\Phi$ are described by rectangular $gl(n-1)$
Young tableaux.
The consideration of gauge symmetries on $(m-1)$-th level goes the same way via identification of Stueckelberg-like contributions
to the transformation law and shows that the reducibility parameters of the reduced theory indeed corresponds to
$H^{-m}$ (\ref{fincoh}).

Noting that $\brst_0$ acts by a linear combination of $\cS^i$ and $\cS^\dagger_i$
we obtain  for $k=0$  component $\varphi$ the following transformation:
\be
\delta \varphi  = \cS^\dagger_i \Lambda^i\;.
\ee
We see that identification $\cS^\dagger_i\equiv D_i$ yields the Labastida gauge law (\ref{gtrLab}).

\vspace{3mm}

The equations of motion that follow from the action (\ref{action}) have the form
\be
\label{eqom}
\brst\, \Psi^{(0)}  = 0\;,
\ee
and are invariant with respect to the gauge transformation (\ref{gaugebrst}). In terms of the components
$\Psi^{(0)}=\Phi+c_0 C$ equations take the form
\be
\label{e1}
\Box \Phi - \brst_0 C = 0\;,
\ee
\be
\label{e2}
\brst_0\Phi + Z_+ C=0\;.
\ee
We observe that all fields $C$ enter the second field equation algebraically and hence can be fully
eliminated by expressing in terms of the first derivatives of fields $\Phi$. Indeed,
similarly to the gauge transformation law analysis the corresponding term in the field equations is expressed
as $Z_+ C$. Noting that fields $C$ are not in the kernel of $Z_+$ we conclude that all of them
can be expressed through the appropriate combinations of $\brst_0\Phi$.

To analyze the field equations for the component $\varphi$ we start with $k=0$ and obtain
\be
\Box \varphi - \cS^\dagger_m C^m=0\;,
\ee
and
\be
\cS^n \varphi -\cS^\dagger_m \Phi^{n|m} -  C^n = 0\;.
\ee
By solving the second equation for the auxiliary field $C^m$ and substituting the result in the
first equation we obtain
$
\Box \varphi - \cS^\dagger_m \cS^m \varphi+\cS^\dagger_m \cS^{\dagger}_n \Phi^{m|n} =0
$.
Taking into account trace relation (\ref{trrel1}) we  finally get the Labastida
field equation
\be
\Big(\Box - \cS^\dagger_m \cS^m+\frac{1}{2}\cS^\dagger_m \cS^{\dagger}_n T^{mn}\Big)\varphi =0\;.
\ee

\section{(Generalized) Poincar\'e modules}
\label{sec:GPM}

\subsection{$Q$-cohomology and the unfolded formulation}\label{sec:Q-coh}

According to the general strategy~\cite{\BGST} given a parent form of the theory,
the unfolded formulation can be obtained reducing to the cohomology of the fiber part of the
BRST operator ~\eqref{parent1}. This is equivalent to reducing the theory \eqref{parent00} to the cohomology of $Q$.
Eliminating the generalized auxiliary fields associated to the contractible pairs
for $Q$ the theory reduces to that determined by the reduced BRST operator of the form~\cite{\BGST}
\begin{equation}
\brst_{unf}=\derham-\tilde\sigma\,,
\end{equation}
where $\derham$ is de Rham differential $\theta^a\dl{x^a}$ and $\tilde \sigma$ is the reduction
of $\sigma=\theta^a\dl{y^a}$ to $Q$-cohomology. In this way one describes
the theory in terms of the fields taking values in $Q$-cohomology only.

In order to explicitly describe the unfolded form of the theory one needs to know $Q$-cohomology.
In the vanishing ghost number it has been already computed in
Section~\bref{sec:Q}. In order to compute $Q$-cohomology at all the remaining ghost numbers, {\it i.e.}
$-(n-1),-(n-2),\ldots,0$, we need some additional algebraic tools.

\subsubsection{$sl(2)$ cohomology}
\label{Sec: sl(2)}

Let $a^A,y^A$ be two sets of variables which we allow to be bosonic or fermionic of the same
Grassmann parity, $\p{a^A}=\p{y^A}$.
On the space of polynomials in $a,y$ we define the $sl(2)$ algebra
\begin{equation}
\label{sl2}
 J=a^A\dl{y^A}\,,
 \qquad \bar J=y^A\dl{a^A}\,,
 \qquad h=\commut{J}{\bar J}=a^A\dl{a^A}-y^A\dl{y^A}\,.
\end{equation}
Extending the space by the ghost variable $b$ with $\gh{b}=-1$ one considers the following operators
\begin{equation}
 q=J\dl{b}\,,\qquad  \bar q=\bar J b\,.
\end{equation}
Both  are obviously nilpotent and act on the space of polynomials $\phi =\phi_1+b\phi_2$.
The operator $q$ has the same structure as $Q$ we are interested in
while $\bar q$ is a kind of anti-BRST operator associated to $q$. The cohomology of both $q$
and $\bar q$ can easily be computed using the $sl(2)$ representation theory. Namely, the representatives
for both $q$ and $\bar q$ can be taken in the form $\phi$ with $\bar J\phi_1=0$ and $J\phi_2=0$.
We see that the $q$ and $\bar q$ cohomology are not only isomorphic but are represented by the
same elements. Moreover, one observes that the cohomology representatives chosen in this way can
be singled out by
\begin{equation}
 q\phi=\bar q\phi=0\,.
\end{equation}

The same is of course true if instead of the space of polynomials
and the algebra \eqref{sl2} one takes an arbitrary representation
space of the $sl(2)$ algebra formed by $J,\bar J,h=\commut{J}{\bar
J}$. The only requirement is that the entire representation space
is decomposable into the direct sum of finite-dimensional
irreducible $sl(2)$-modules. In particular, if among variables
$a^A,y^A$ there is a fermionic pair $\alpha,\gamma$ then the
statement is also true if instead of $\alpha\dl{\gamma}$ and
$\gamma\dl{\alpha}$ terms in $J,\bar J$ one takes $\alpha\gamma$
and $\dl{\gamma}\dl{\alpha}$, respectively. This is because
$\alpha\gamma$, $\dl{\gamma}\dl{\alpha}$ and
$\commut{\alpha\gamma}{\dl{\gamma}\dl{\alpha}}=\alpha\dl{\alpha}+\gamma\dl{\gamma}-1$
also form $sl(2)$.

\subsubsection{$Q$-cohomology}

We now turn to the computation of the $Q$-cohomology in the subspace \eqref{part-Weyl}.
As we have seen the representatives of $Q$-cohomology at zeroth ghost degree  can be chosen to be annihilated by
the upper-triangular subalgebra
\begin{equation}
\label{U0}
U_0 = \Big\{N_i{}^j\;\;\;\;\; i>j\;,
\quad
\bsd{}^i\Big\}\;,
\end{equation}
of the $sl(n)$ algebra generated by $N_i{}^j\,\,\, i\neq j$, $\sd_i$, $\bsd{}^i$.

In fact there are other choices for the upper-triangular subalgebra
containing  $N_i^j\,\,\, i>j$. More precisely, there are $n$ subalgebras
\be
U_p = \Big\{N_i{}^j\;\;\;\;\; i>j\;,
\quad
\bsd{}^i\;,\;\;i=1,...,n-p-1\;,
\quad
\sd_j\;,\;\; j=n-p,..., n-1\Big\}\;,
\ee
which are upper-triangular and contain $N^j_i\,\,\, i>j$.
For $p=0$ this indeed gives \eqref{U0}.

Each $U_p$ define a Poincar\'e module $\modM_p$ (one can consistently define the Poincar\'e module structure in the same way as for $\modM_0$).
The conditions read explicitly as
\be
\label{Gauge-module1}
 T^{IJ}\phi=0 \,,\qquad  N_i{}^j\phi=0\,\,\,\;\; i>j\,,
\ee
\be
\ba{l}
\bsd{}^i\phi=0\;,\qquad N_i\phi=s_i\,\phi\;,\quad\;\;\;\;\;\;i=1,\ldots,n-p-1\;,
\\
\\
\sd_j\phi=0\;, \qquad\; N_j\phi=(s_j-1)\phi\;, \quad \;\;j=n-p,\ldots, n-1\;,
\ea
\ee
and can be represented by Young tableaux of the form
\be
\label{YT3}
\begin{picture}(75,70)(0,2)
\multiframe(0,-11.1)(10.5,0){1}(10,10){}\put(14,-10){$s_1$}
\multiframe(0,-1)(10.5,0){2}(10,10){}{}\put(25,2){$s_{2}$}
\put(-0.2,0){\line(0,1){45}}
\multiframe(0,20)(10.5,0){3}(10,10){}{}{}{}{} \put(38,22.5){$s_{0}$}%\put(-70,22.5){${\small (l+1)^{th}\; raw}$}
\put(3,35){$.\;.\;.\;.$}
\put(3,15){$.\;.\;.$}

%\multiframe(0,30.5)(10.5,0){2}(10,10){}{}\put(25,30.5){$s_{0}$}
\multiframe(0,41.5)(10.5,0){5}(10,10){}{}{}{}{} \put(60,42.5){$s_{n-2}-1$}
\multiframe(0,52)(10.5,0){7}(10,10){}{}{}{}{}{}{} \put(79,55){$s_{n-1}-1$}
\end{picture}
\ee

\vspace{.2cm}

\noindent with the weight $s_0$ such that
\be
\label{YT4}
s_{n-p-1}\leq s_0\leq s_{n-p}-1\;.
\ee
From the above inequality it follows that if $s_i=s_{i-1}$ then module $\modM_{n-i}$ is empty.
Note that whatever weights $s_i$ are module $\modM_0$ is always nonempty.
Let us also note that $\modM_{n-1}$ coincides with the module defined by
\eqref{Gauge-module}, \eqref{Gauge-module-2} if one shifts weights $s_i\to s_i-1$.

We have the following
\begin{prop} The cohomology of $Q$ in subspace \eqref{part-Weyl} at ghost degree $p$ can
be identified with the subspace $\modM_p$. In particular, $\modM_p$ is naturally a Poincar\'e module for any $p$.
\end{prop}
The second statement immediately follows from $\commut{Q}{M_{ab}}=\commut{Q}{P_a}=0$.
The proof of the Proposition is given in Appendix  \bref{sec:B}. In what follows
we explicitly demonstrate the computation of $Q$-cohomology for the first nontrivial
case $n=3$.

%\subsubsection{The case of $n=3$}
 The cohomology of the BRST operator $Q=\sd_1\dl{b_1}+\sd_2\dl{b_2}$ in the
subspace
\eqref{part-Weyl} can be identified with the cohomology of
\begin{equation}
 \hat Q_0=\chi N_2{}^1+\sd_1\dl{b_1}+\sd_2 \dl{b_2}+\chi b_2\dl{b_1}\equiv
\chi N_2{}^1+\sd_2 \dl{b_2}+(\sd_1+\chi b_2)\dl{b_1}
\end{equation}
evaluated in the space of elements satisfying $T^{IJ}\phi=0$ along with the weight conditions and represented by
$\chi$-independent elements. Here we have introduced ghost variable $\chi$
associated to the constraint $N_2{}^1$ that generates the cubic ghost term in $\hat Q_0$. Indeed,
for a $\chi$-independent element the cocycle condition implies $\cN_2{}^1\phi=0$.

Along with $\hat Q_0$ let us consider another nilpotent operator
\begin{equation}
 \hat Q_1=\chi N_2^1+\bsd{}^1b_1+\sd_2 \dl{b_2}+\dl{b_2}\dl{\chi}{b_1}=
\chi N_2^1+\sd_2 \dl{b_2}+(\bsd{}^1+\dl{b_2}\dl{\chi})b_1\,.
\end{equation}
This can be seen as a BRST operator implementing the conditions from upper-triangular subalgebra $U_1$ if one
flips the ghost number assignment for the variable $b_1$.
The difference
between $\hat Q_0$ and $\hat Q_1$ is in
\be
q_1=(\sd_1+\chi b_2)\dl{b_1}
\ee
replaced by
\be
\bar q_1=(\bsd{}^1+\dl{b_2}\dl{\chi})b_1\;.
\ee
As suggested by the notations these two operators are indeed particular cases of $q$ and $\bar q$ discussed above for
$a^A=\{a_1^a,\chi\}$ and $y^A=\{y^a,b_2\}$.

In fact, cohomology of $\hat Q_0$ and $\hat Q_1$ are identical. To see this let us reduce both
cohomological problems
to the cohomology of $q_1$ and $\bar q_1$, respectively (this can be achieved decomposing the operators
in the homogeneity degree in $b_1$). Choosing as representatives the subspace $q_1\phi=\bar q_1\phi=0$
one observes that $\hat Q_0$ and $\hat Q_1$ act in this subspace.
Moreover, in this subspace they simply coincide. This proves that cohomology of $\hat Q_0$ and
$\hat Q_1$ are isomorphic and the representatives
can be taken the same.

In exactly the same way one proves that the cohomology of $\hat Q_1$ is identical to the cohomology of $\hat Q_2$ given by
\begin{equation}
 \hat Q_2=\chi N_2^1+\bsd{}^1b_1+(\bsd{}^2+\chi\dl{b_1})b_2\,,
\end{equation}
which can be considered a BRST operator implementing the conditions from $U_2$ if one in addition
changes the ghost number assignment for
$b_2$. Analogously, the difference
between $\hat Q_1$ and $\hat Q_2$ is in
\be
q_2=(\sd_2-b_1\dl{\chi})\dl{b_2}
\ee
replaced by
\be
\bar q_2=(\bsd{}^2-\chi\dl{b_1})b_2\;.
\ee
Once again, above operators are particular cases of $q$ and $\bar q$
with  $a^A=\{a_2^a, -b_1\}$ and $y^A=\{y^a,\chi\}$.

Operators $\hat Q_0, \hat Q_1, \hat Q_2$ allow us to immediately compute the cohomology on the representation
space of elements $\phi = \phi^0 + b_1\phi^1+b^2\phi_2 + b_1b_2\phi^{12}$.
In particular, for elements whose
representatives have the form $b_1\phi^1+b^2\phi_2 $ the cocycle conditions with respect to $\hat Q_0$ and $\hat Q_1$
imply $\phi_1=0$ and
\begin{equation}\label{classes-2}
 N_2{}^1\phi^2=0\,, \quad \bsd{}^1\phi^2=\sd_2\phi^2=0\,,
\end{equation}
{\it i.e.} the conditions for $\modM_1$.
For elements of the form $b_1b_2\phi^{12}$ the cocycle condition with respect to $\hat Q_0$ gives
\begin{equation}
 N_2{}^1\phi^{12}=0\,, \quad \bsd{}^1\phi^{12}=\bsd{}^2\phi^{12}=0\,,
\end{equation}
{\it i.e.} the conditions for $\modM_2$. Finally, for ghost-independent $\phi^0$ the cocycle conditions with respect to $\hat Q_2$
give the conditions identified in Section \bref{sec:Q}, {\it i.e.}
\begin{equation}
 N_2{}^1\phi^0=0\,, \quad \bsd{}^1\phi^0=\bsd{}^2\phi^0=0\,.
\end{equation}

The above consideration results in the observation that
any cohomology class has (in fact, a unique) representative satisfying
\be
\hat Q_0\phi=\hat Q_1\phi=\hat Q_2\phi=0\;.
\ee
Let us now recall that in addition $\phi$ satisfies the tracelessness condition $T^{IJ}\phi=0$,
and Young symmetry and the weight conditions
$N_i^j\phi=0\,\,\, i>j\;,\;$ $\hat\cN_i \phi=s_i\,\phi$. Below we
describe all the solutions to these conditions.

\begin{itemize}

\item
Module $\modM_0$ singled out by
$
N_2 \phi^0 = s_2\, \phi^0,\; N_1 \phi^0 = s_1\, \phi^0
$
is described by the following Young tableaux
\be
\label{YTM0}
\begin{picture}(120,70)(0,2)
\put(-50,42.5){$\modM_0\;:$}

\multiframe(0,31.0)(10.5,0){4}(10,10){}{}{}{}\put(45,32){$s_1$}
\multiframe(0,41.1)(10.5,0){5}(10,10){}{}{}{}{} \put(60,42.5){$s_2$}
\multiframe(0,52)(10.5,0){10}(10,10){}{}{}{}{}{}{}{}{}{} \put(110,55){$s_0$}

\put(130,42.5){$,\qquad s_0\geq s_2\;.$}

\end{picture}
\ee

\item \noindent Module $\modM_1$ singled out by
$
N_2 \phi^2 = (s_2-1) \phi^2, \;N_1 \phi^2 = s_1\, \phi^{2}
$
is described by the following Young tableaux
\be
\label{YTM1}
\begin{picture}(120,70)(0,2)
\put(-50,42.5){$\modM_1\;:$}

\multiframe(0,31.0)(10.5,0){4}(10,10){}{}{}{}\put(45,32){$s_1$}
\multiframe(0,41.1)(10.5,0){5}(10,10){}{}{}{}{} \put(60,42.5){$s_0$}
\multiframe(0,52)(10.5,0){7}(10,10){}{}{}{}{}{}{} \put(80,55){$s_2-1$}

\put(110,42.5){$,\qquad s_1\leq s_0\leq s_2-1\;.$}

\end{picture}
\ee
\item \noindent Module $\modM_2$ singled out by
$N_2 \phi^{12} = (s_2-1) \phi^{12}\;, N_1 \phi^{12} = (s_1-1) \phi^{12}$
is described by the following Young tableaux
\be
\label{YTM2}
\begin{picture}(120,70)(0,2)
\put(-50,42.5){$\modM_2\;:$}

\multiframe(0,31.0)(10.5,0){4}(10,10){}{}{}{}\put(45,32){$s_0$}
\multiframe(0,41.1)(10.5,0){5}(10,10){}{}{}{}{} \put(60,42.5){$s_1-1$}
\multiframe(0,52)(10.5,0){7}(10,10){}{}{}{}{}{}{} \put(80,55){$s_2-1$}

\put(110,42.5){$,\qquad 0\leq s_0\leq s_1-1\;.$}

\end{picture}
\ee

\end{itemize}

In the case of coinciding weights $s_1=s_2=s$ one gets $N_1\phi^2=s\,\phi^2$ and
$N_2\phi^2=(s-1)\phi^2$. But these contradict $N_2{}^1\phi^2=0$ because
it implies that the number of $a_2$ is greater or equal than that of $a_1$. It follows
that module $\modM_1$ is empty and the remaining modules are described
by
\be
\begin{picture}(90,70)(0,2)
\put(-50,42.5){$\modM_0\;:$}

\multiframe(0,31.0)(10.5,0){8}(10,10){}{}{}{}{}{}{}{}\put(88,32){$s$}
\multiframe(0,41.1)(10.5,0){8}(10,10){}{}{}{}{}{}{}{} \put(88,42.5){$s$}
\multiframe(0,52)(10.5,0){10}(10,10){}{}{}{}{}{}{}{}{}{} \put(110,55){$s_0$}

\put(130,42.5){$,\qquad s_0\geq s\;.$}

\end{picture}
\ee
\be
\begin{picture}(90,70)(0,2)
\put(-50,42.5){$\modM_2\;:$}

\multiframe(0,31.0)(10.5,0){4}(10,10){}{}{}{}\put(45,32){$s_0$}
\multiframe(0,41.1)(10.5,0){7}(10,10){}{}{}{}{} {}{}\put(80,42.5){$s-1$}
\multiframe(0,52)(10.5,0){7}(10,10){}{}{}{}{}{}{} \put(80,55){$s-1$}

\put(110,42.5){$,\qquad 0\leq s_0\leq s-1\;.$}

\end{picture}
\ee
\noindent
The discussed above $\modM_0$, $\modM_1$, and $\modM_2$ can be recognized as modules appearing
within  the unfolded
formulation for two-row fields \cite{Skvortsov:2008vs}. More precisely, $p$-form fields
with $p=0,1,2$ of the unfolded approach take values in $\modM_p$.
The case $s_1=s_2$ was also considered in \cite{Alkalaev:2003qv,Alkalaev:2003hc}.
A detailed discussion of the relationship with the unfolded formulation for any $n$ is given
in the next Section.

%\newpage
\subsubsection{Unfolded formulation}
\label{sec: unfolding}

Any cohomology class of the form $b_{n-1}b_{n-2}\ldots
b_{n-p}\phi_p$ gives rise to the ghost-number-zero element of the
form $\theta^{a_1}\ldots\theta^{a_p}\,b_{n-1}b_{n-2}\ldots
b_{n-p}\,\phi_{a_1\ldots\, a_p}$. These in turn give rise to the
physical fields that are $p$-forms on Minkowski space. One then
finds that the space of physical fields of the theory
\eqref{parent00} reduced to $Q$-cohomology is given by differential
forms of degrees $0,1,\ldots ,n-1$ taking values in respectively the
cohomology spaces at ghost number $0,-1,\ldots,-n+1$ described by
Lorentz Young tableaux \eqref{YT3}, (\ref{YT4}). If $s_l=s_{l-1}$
then the $(n-l)$-form is missing so that remaining fields correspond
to the rectangular blocks of the Young tableaux with rows of the
length $s_l$. One then concludes that the spectrum of unfolded
fields coincide with that proposed in \cite{Skvortsov:2008vs}.

In order to identify the unfolded equations and gauge symmetries
one is to find a reduced BRST operator $\tilde{\brst}$ acting in
the $Q$-cohomology.  More precisely, the reduced operator have the
form $\tilde\brst=\derham-\tilde\sigma$ where $\tilde\sigma$ is
the differential $\sigma=\theta^a \dl{y^a}$ reduced to the
$Q$-cohomology. We also save notation $\derham$ for the restriction
of $\derham=\theta^a\dl{x^a}$ to the $Q$-cohomology. In order to compute $\tilde\sigma$ we follow the
procedure of \cite{\BGST}. To this end we introduce minus the target-space ghost number
as an additional grading. Then the entire
representation space $\cH$, {\it i.e.} the space \eqref{part-Weyl}
tensored with the Grassmann algebra in $\theta^a$ is decomposed into the direct
sum $\cH=\cE\oplus\cG\oplus\cF$, where $\cE$ is the subspace of
representatives of the $Q$-cohomology, $\cG=\im Q$, and $\cF$ the
complementary subspace. $Q$ determines the invertible map from
$\cF$ to $\cG$. Let also $\rho : \cG \to \cF$ be the inverse to
$Q$, {\it i.e.} $Q\rho g=g$ for any $g\in\cG$. It follows that operator $\rho$ can be chosen
to have
a degree $+1$ with respect to ghosts $b_i$ and variables $y^a$, and
a degree $-1$ with respect to variables $a^a_i$.
We also assume that
$\rho$ is extended to $\Ker Q=\cE\oplus \cG$ such that $\rho e=0$
for any $e \in \cE$. Given such $\rho$ the expression for
$\tilde\sigma:\cE\to\cE$ reads as~\cite{\BGST}
\begin{equation}
\label{t-sigma}
 \tilde\sigma=\Pi_\cE(\sigma -(\sigma \rho)\, \sigma+(\sigma \rho)\, (\sigma \rho)\, \sigma-\ldots)\,,
\end{equation}
where $\Pi_\cE$ denotes the projector to $\cE$. If $\cE$ contains
cohomology classes with ghost numbers from $0$ to $n-1$ then in
general only first $n$ terms can be non-vanishing in this series.
Also in $d$-dimensions the $(d+1)$-st term necessarily vanish but
this does not play a role because $n\leq [\frac{d}{2}]$.

Let us first make some general observations on the explicit
structure of $\tilde\sigma$. Let $f_i \in \cE$, $i=0,\ldots,n-1$
has the form $f_i=b_{i+1}\ldots b_{n-1} \phi_i\in\modM_{n-i-1}$ (it is assumed
that $f_{n-1}=\phi_{n-1}$ and $\phi_i$ depends also on $\theta^a$).
Then the term $\Pi_\cE
(\sigma\rho)^l\sigma f_i$ with $l\geq 1$ in $\tilde\sigma f_i$ can
be nonvanishing only if $\#y=s_{i}=s_{i-1}=\ldots=s_{i-l+1}$,
where $\#y$ equals to $s_0$ and denotes the homogeneity degree of $f_i$ in
variables $y^a$. This can be easily seen by counting the number of
ghosts and oscillators and then comparing with the structure of
the $Q$-cohomology. Moreover,  this is the only nonvanishing terms in
the whole series for $\tilde\sigma f_i$ (if $s_0\neq s_i$ then the
only nonvanishing term is $\Pi_\cE\sigma$) provided $s_{i-l+1}\neq
s_{i-1}$. This implies that for a given $f_i$ the only term that
contributes corresponds to the rectangular block with the upper
row representing $a^a_i$ of the Young tableau encoding the
symmetry properties of $f_i$.

Let now $f_i\in \cE$ be of the form above and such that in addition
\begin{equation}
 \sd_i f_i=\bar S^{\dagger i} f_i=0\,,\qquad \ldots\,,  \qquad  \sd_{i-l} f_i=\bar S^{\dagger i-l}f_i=0\,.
\end{equation}
This means that the Young tableaux representing $f_i$ contains the
rectangular block of the height $l+1$ (with the rows corresponding
to $y,a_i,\ldots, a_{i-1}$). Let us consider $Q(\bar S^{\dagger i}
b_i)\sigma f_i$. Among all the operators $\dps\sd_m\dl{b_m}$ entering
$Q$ those with $m<i$ act trivially because $f_i$ does not depend
on $b_m$ with $m<i$ while those with $m>i$ commute with $\bar
S^{\dagger i}$ in the subspace of elements satisfying the Young
symmetry conditions $N_i{}^j\phi=0\,\,\, i>j$. Moreover
$\dps\sd_m\dl{b_m}\sigma f_i=0$ for $m>i$ because of the cocycle
condition $Q\sigma f_i=0$. The remaining term $\sd_i\dl{b_i}$
gives $\commut{\sd_i}{\bar S^{\dagger i}}\sigma f_i-\bar
S^{\dagger i} \sigma \sd_i f_i$. The second term vanishes because
$\commut{\sd_i}{\sigma}=0$ and $\sd_i f_i=0$ according to the
assumption. The first term gives $(a_i\dl{a_i}-y\dl{y})\sigma
f_i$. Because for $f_i$ one has $(a_i\dl{a_i}-y\dl{y})f_i=0$ one
gets $(a_i\dl{a_i}-y\dl{y})\sigma f_i=\sigma f_i$ so that
\begin{equation}
 Q(\bar S^{\dagger i} b_i)\sigma f_i=\sigma f_i\,.
\end{equation}
This shows that for an element of the form $\sigma f_i$ one can
consistently define $\rho$ according to  $\rho\sigma f_i = \bar
S^{\dagger i} b_i \sigma f_i$. Using $\bar S^{\dagger i} b_i
\sigma f_i=\bar\sigma^i b_i f_i$ where $\bar\sigma^i=\theta^a
\dl{a_i^a}$ it is easy to see that $f_{i-1}=\bar S^{\dagger i} b_i
\sigma f_i$ again  satisfies $Q\sigma f_{i-1}=0$ and
\begin{equation}
\sd_{i-1} f_{i-1}=\bar S^{\dagger i-1}f_{i-1}=0,
\qquad  \ldots\,, \qquad
\sd_{i-l} f_i=\bar S^{\dagger i-l}f_i=0\,,
\end{equation}
so that the construction can be iterated defining the action of $\rho$ in all the nonvanishing
terms in \eqref{t-sigma}. For instance in the setting above one gets
\begin{equation}
 (\sigma \rho)^l\sigma f_i= \sigma\bar\sigma^{i-l}b_{i-l}\bar\sigma^{i-l+1}b_{i-l+1}\ldots\bar\sigma^{i}b_{i}f_i\,.
\end{equation}
It turns out that using the fact that all the other terms in \eqref{t-sigma} for a particular $f_i$ vanish one
can write a closed expression for $\tilde\sigma$
\begin{equation}
 \tilde\sigma=\Pi_\cE(\sigma-\sum_{i=1}^{n-1}\sigma\bar\sigma^ib_i+\sum_{i<j} \sigma\bar\sigma^ ib_i\bar\sigma^jb_j-\ldots)\,.
\end{equation}
For example for $n=3$ one gets
\begin{equation}
 \tilde\sigma=\Pi_\cE(\sigma-\sigma\bar\sigma^1 b_1-\sigma\bar\sigma^2 b_2+ \sigma\bar\sigma^1b_1\bar\sigma^2 b_2)\,.
\end{equation}
Note that if $s_1=s_2$ only the first and the last terms contribute because the cohomology class $\modM_1$
is missing in this case.

\subsection{Generalized Poincar\'e module}\label{sec:gen-brst}

As we have seen the spectrum of the unfolded fields can be
described as a zero-ghost-number $Q$-cohomology $\modM$ evaluated
in the space \eqref{part-Weyl} tensored  with the Grassmann
algebra in $\theta^a$. This space is graded by the
homogeneity in $b_i$ (this degree is known in the literature as the target space
ghost number) and its zeroth
degree component coincides with $\modM_0$ while the higher degree
components are $\modM_p$ tensored with the  $p$-th homogeneous
subspace of the Grassmann algebra in $\theta^a$. In the gauge
description of the model this space replaces the starting point
module $\modM_0$ entering the gauge invariant description
\eqref{covconst}. In fact it is easy to see that the unfolded
equations of motion for $b_i$-independent fields indeed reproduce
\eqref{covconst-P} for $\modM_0$-valued $0$-form. This suggests that
$\modM$ is a natural generalization (extension) of $\modM_0$
referred to in what follows as the Generalized Poincar\'e module.
Note that from
the BRST theory viewpoint it can be natural to consider
all the $Q$-cohomology (not only at zeroth ghost degree). This
can be seen as a BRST extension of $\modM$. Along with the fields of
the unfolded formulation it contains all the respective ghost fields and antifields.

Let us briefly discuss the Poincar\'e module structure of $\modM$. To identify this structure it is convenient to
use the alternative realization \eqref{bar-Poincare} of the Poincar\'e algebra in the theory determined by \eqref{parent0}.
In reducing the intermediate formulation \eqref{parent00} to
$Q$-cohomology the generators (\ref{bar-Poincare}) are also
reduced to some operators acting on $Q$-cohomology valued fields.
Because Lorentz generators $\bar M_{ab}$ maps representatives
(\ref{classes}) to themselves and therefore their reduction is
given by the same formulas. To obtain a reduction of $\bar P_a$
one should be more careful. The form of the reduced operator can
be computed using, {\it e.g.}, the formulas from~\cite{Barnich:2005ga}. It
turns out, however, that this can equivalently be inferred from
the $\tilde\sigma$ through $\tilde\sigma=\theta^a \tilde P_a$.
Indeed, because $\theta$-variables enter the $Q$-cohomology through the tensor factor reducing $\sigma$ to
$Q$-cohomology is the same as reducing $P_a$ to $Q$-cohomology and
then constructing $\tilde\sigma=\theta^a \tilde P_a$.

Inspecting the explicit form of the reduced generators one finds
that they also define the Poincar\'e module structure on $\modM$.
This can be seen by, {\it e.g.}, identifying $\modM$ with constant
$\modM$-valued fields. As it follows from the explicit form of
$\tilde \sigma$ generator $\tilde P_a$ (in contrast to the Lorentz
generators that are unchanged) does act between
different degree components of $\modM$. From this point of view
$\modM$ can be thought as modules $\modM_p$ tensored with the
algebra of $\theta^a$ and nontrivially glued together.

Remarkably, $\modM$ can be equipped with two in general different
Poincar\'e module structures. The one determined by $P_a$ (the
operator $\dl{y^a}$ acting in the cohomology of $Q$), for which
the generalized Poincar\'e module is a direct sum of Poincar\'e
modules appearing in different degrees and another one determined
by $\tilde P$ (the reduction of $\dl{y^a}$ to the unfolded formulation)
for which the generalized Poincar\'e module is not a direct sum in
general. The tricky point here is that in both cases one reduces
$\dl{y^a}$ to the $Q$-cohomology but the form of the reduced
operator depends on the total BRST operator. In the algebraic
setting of Section~\bref{sec:Q} this total operator is $Q$ itself
while for $\tilde P$ the total BRST operator is $\hat
\brst=\derham-\sigma+Q$.

As a final remark note that one can also define two different theories
determined by the BRST operators $\derham-\tilde\sigma$  and
$\derham-\theta^aP_a$. While the first one is the genuine gauge
theory the second one is the direct sum of the gauge-invariant
theory \eqref{covconst} and a bunch of the decoupled topological
theories for differential forms of nonzero degrees.

\subsection{Wigner approach}
\label{sec: Wigner}

Another Poincar\'e module associated to the starting point module
$\modM_0$ (Weyl module) can be identified by considering the space
of gauge inequivalent solutions of the theory in the appropriate
functional space. To construct this module explicitly we use the
standard BRST first-quantized description of the theory
constructed in Section \bref{sec:standard}. Recall that the theory
is determined by the BRST operator $\brst$ given by
\eqref{standard-brst} and the representation space is formed by
$y^a,\theta^a$-independent elements satisfying the BRST extended
trace, Young symmetry, and the weight conditions
\eqref{brstextconst}.

Let us now describe the space of gauge inequivalent configurations
of the theory in the space of functions where $\dl{x^a}$ act
diagonally ({\it i.e.} in the momentum representation). This space of
inequivalent configurations can be identified with the
zero-ghost-number $\brst$-cohomology. Because $\brst$
commutes with $\dl{x^a}$ it is enough to compute cohomology in a
momentum eigenspace where $\dl{x^a}\phi=p_a\phi$. Assuming
$p_a\neq 0$ (and hence disregarding the so-called zero-momentum
cohomology) the cohomology can be easily computed using the
arguments similar to the standard light-cone gauge. We follow
\cite{Barnich:2005ga} (see also \cite{Henneaux:1987cpbis} for a more
traditional approach) where this computation has
been explicitly carried over in the similar terms for $n=2$.

Let us introduce the light-cone components $+,-,\alpha$ of the momenta and the oscillators.
Assuming $p^+\neq 0$  consider the following degree in the representation space
\begin{equation}
  \begin{gathered}
   \deg(a_i^+)=2, \qquad  \deg(a^-_i)=-2,\\
    \deg(c_i)=1,  \qquad \deg(b_i)=-1\,,
\end{gathered}
\end{equation}
with all the other variables carrying vanishing degree \cite{Aisaka:2004ga, Kato:1982im}.
The BRST operator \eqref{standard-brst} can be expanded into the components of definite degree
as $\brst=\brst_{-1}+\brst_{0}+\brst_{1}+\brst_2$. The lowest degree component of $\brst$ reads as
\begin{equation}
\label{lcder}
 \brst_{-1}=p^+(c_i \dl{a^+_i}+{a^-_i} \dl{b_i})\,
\end{equation}
and can be seen as a version of de Rham differential multiplied by $p^+$.
Because the degree is bounded from below (in the space of polynomials in oscillators $a_i$) one can first reduce the problem to the cohomology of $\brst_{-1}$ in the
subspace singled out by the conditions~\eqref{brstextconst} (this is consistent
as~\eqref{brstextconst} commute with $\brst_{-1}$).

It turns out that this cohomology can be obtained by restricting
the $\brst_{-1}$-cohomology evaluated in the space of all
polynomials to the subspace~\eqref{brstextconst}. This is obvious
for the constraints $\cN_i-s_i$ because $\brst_{-1}$ do not mix
different eigenspaces. As for the remaining constraints $\cT^{ij}$
and $\cN_i{}^j\,\,, i>j$ they can be added as the additional
constraints to BRST operator $\brst_{-1}$ with their own ghost
variables $\xi_A$ so that the required cohomology can be identified
with cohomology of the extended BRST operator whose
representatives can be chosen $\xi_A$-independent. The extended
BRST operator has the structure $\brst^\prime=\brst_{-1}+\xi_A
\cT^A+\text{ghost terms}$, where $\cT^A$ is a collective notation
for the constraints $\cT^{ij}$ and $\cN_i{}^j\,\,, i>j$. Observing
that the constraints $\cT^A$ carry vanishing degree one reduces
the cohomological problem for $\brst^\prime$ to the cohomology of
$\brst_{-1}$. In the space of all polynomials
$\brst_{-1}$-cohomology is given by a subspace $\cE$ of
$c_i,b_i,a_i^+,a^-_i$-independent elements. The reduced BRST
operator has the form $\xi_A \tilde\cT^A+\text{ghost terms}$,
where the reduced constraints $\tilde \cT^A$ can be shown to be
just original constraints $\cT^A$ restricted to $\cE$. For a
$\xi_A$-independent element from $\cE$ the cocycle condition imply
\begin{equation}
\label{wig-cond}
\begin{gathered}
 \tilde T^{ij}\phi=\eta^{\alpha\beta}\dl{a_i^\alpha}\dl{a_j^\beta}\phi=0\,,\qquad
\tilde N_i{}^j\phi=a_i^\alpha\dl{a_j^\alpha}\phi=0 \,\,\, i>j\,,\\
\tilde N_i=a_i^\alpha\dl{a_j^\alpha}\phi=s_i\phi\,.
\end{gathered}
\end{equation}
This gives an explicit description of $\brst_{-1}$-cohomology in the
subspace~\eqref{brstextconst}.

The reduced theory is then determined by the reduced BRST operator
\begin{equation}
\label{lc-brst}
 \tilde\brst=c_0 (p^ip_i-2p^+p^-)\,,
\end{equation}
defined on the subspace \eqref{wig-cond}. Note that $\brst_{1}$ and $\brst_2$
do not contribute because the cohomology is concentrated in zeroth degree.
The zero-ghost-number
$\tilde\brst$-cohomology in the momenta eigenspace is given by an arbitrary
$c_0$-independent elements satisfying \eqref{wig-cond} multiplied
by $\delta(p^2)$. Finally, the cohomology can be identified with
the ``Wigner module'', {\it i.e.} functions on the mass-shell $p^2=0$
with values in the subspace \eqref{wig-cond}. One can speculate
that the procedure above establishes an explicit duality transform
between the Weyl module (zero-ghost-number cohomology of $\brst$
in the space of polynomials in $x^a$) and the Wigner
module (zero-ghost-number cohomology of $\brst$ in the space
where $\dl{x^a}$ is diagonalizable).
Note that because $\brst_{-1}$ is symmetric with respect to the inner product the reduction is consistent with the inner product.

The reduced action can be readily obtained in the form (see \cite{Barnich:2005ga} for details)
\begin{equation}
 S^{lc}=\half\int d^dp\, dc_0\, \inner{\tilde\Psi^{(0)}} {\tilde\brst \tilde\Psi^{(0)}}_0\,,
\end{equation}
where the field $\Psi^{(0)}$ now takes values in the subspace
\eqref{wig-cond} and $\inner{\cdot}{\cdot}_0$ denotes the Fock
space inner product (see \eqref{innerprod}) restricted to the
subspace generated by the transversal oscillators. This is indeed
the standard light-cone action for the transversal degrees of
freedom. As it should be there is no leftover gauge symmetry. It
is easy to see that conditions \eqref{wig-cond} are the
irreducibility conditions for the $so(d-2)$ which is a Lie algebra
of Wigner little group. One then concludes that the transversal
degrees of freedom form an irreducible representation of $so(d-2)$
determined by weights $s_{n-1}\geq s_{n-2}\geq    \ldots \geq
s_1$. Note also that these conditions form an upper triangular
subalgebra of $sp(2n-2)$ that is Howe dual to $so(d-2)$ on the
Fock space of transversal oscillators $a^\alpha_i$. Let us stress
that contrary to the computation of the cohomology in the space of
polynomials in variables $y$ there are no cohomology classes
depending on $c_i,b_i$. In particular, the states analogous to
those in the gauge modules do not appear. This is also due to the
assumption that $p^+\neq 0$. That is why the states from the gauge
modules are often called zero momentum cohomology. These states
are ignored in the Wigner approach.

Because the procedure just described follows the standard steps of the Wigner description of the
unitary irreps one concludes that the gauge theory determined by \eqref{standard-brst} along with
the trace, Young symmetry, and the weight conditions indeed describe a
unitary irrep of the Poincar\'e group in the sense of Wigner approach.

\section{Conclusions and outlooks}
\label{sec:conc}

The above study could clearly be extended in various directions. A
rather natural generalization is to allow for non-vanishing
cosmological constant that implies the $(A)dS_d$ background
geometry. For totally symmetric fields the corresponding parent
formulation was developed in \cite{Barnich:2006pc} and its
extension to the case of arbitrary symmetry type will be
considered elsewhere.

Our formulation can be also generalized
to describe massive fields of any
symmetry type on  Minkowski space. This could be done
using a standard dimesional reduction $d+1\rightarrow d$
thereby obtaining massive field dynamics in $d$ dimensional Minkowski space.
This procedure can be implemented in the BRST theory terms~~\cite{Burdik:2000kj,Bekaert:2003uc} and hence
is directly applicable to the present formulation.

%Alternative approaches to massive field dynamics were developed
%in \cite{Singh,Zin,Buch,Artsukevich,Metsaev, ALL}.

An interesting topic is to develop supersymmetric extensions which
assume an appropriate inclusion of fermionic mixed-symmetry
fields. Within our approach addressing the problem seems to be
straightforward  and reduces to introducing spin-tensors in an
appropriate fashion. This can be  achieved either by considering
polynomials with coefficients in spinorial modules as,
\textit{e.g.}, in~\cite{Metsaev:1995re,Buchbinder:2004gp,Moshin:2007jt} or by
introducing additional oscillators transforming as $so(1,d-1)$
spinors \cite{Vasiliev:1987tk}. Both ways are equivalent and leave
intact the main ingredients of our construction.

Another possible extension has to do with describing dual
formulations (see, \textit{e.g.},~\cite{Boulanger:2003vs,Matveev:2004ac}, and
references therein) of the mixed symmetry fields. These can be
expected to arise through the different realizations of the
Poincar\'e translations. Much less trivial seems the possibility
to give a realization of the same module in the space-time of
different geometry and/or dimension in the spirit of
\cite{Vasiliev:2001dc,Vasiliev:2003ar}.

A natural question that can be asked using the formulation
developed in the paper is whether there exist a mixed symmetry
counterparts of the well known higher spin algebras. Although in
the case of symmetric fields a consistent HS algebra exists only
on AdS space, at the off-shell level one can identify the
analogous structure also in the Minkowski space. Moreover, in the
symmetric field case a natural
framework~\cite{Vasiliev:2005zu,Grigoriev:2006tt} to study this
structure is provided by a version of the intermediate formulation
\eqref{parent00}. From this perspective, the approach developed in
the paper can be a natural tool to study candidate HS algebras for
mixed symmetry fields that in turn can be a first step towards
constructing consistent interactions for mixed-symmetry fields.

\subsection*{Acknowledgments}

We are grateful to G.~Barnich, N.~Boulanger, R.~Metsaev, E.~Skvortsov, M.~Vasiliev for many illuminating  discussions.
This work is supports by the LSS grant Nr 1615.2008.2.
The work of KA is supported in part by grants
RFBR grant Nr 08-02-00963 and the Alexander von Humboldt Foundation grant PHYS0167. The work of
MG is supported by the RFBR grant Nr 08-01-00737 and the Dynasty foundation.
The work of IYuT was supported in part by the RFBR
grant Nr 08-02-01118 and the Dynasty foundation.

%\newpage
\appendix
\section{Structure of the polynomial $sp(2n)$ modules }\label{sec:A}

We choose Chevalley generators in the form
\be
\ba{l}
E_I=T_{I+1}{}^{I}\\
H_I=T_{I+1}{}^{I+1}-T_{I}{}^{I}\\
F_I=T_I{}^{I+1}
\ea
\mbox{for $0\leq I\leq n-2$ and}
\ba{l}
E_{n-1}=T^{n-1\,n-1}\\
H_{n-1}=-T_{n-1}{}^{n-1}\\
F_{n-1}=-\frac{1}{4}T_{n-1\,n-1}
\ea
\ee
We choose orthonormal basis $h_I=-T_{I}{}^{I}$, $0\leq I\leq n-1$ in
the Cartan subalgebra.
Then $H_I=h_I-h_{I+1}$ for $0\leq I\leq n-2$ and $H_{n-1}=h_{n-1}$.
The dual basis $\epsilon_I$ in the space dual to the Cartan subalgebra satisfy
$\inner{\epsilon_I}{h_J}=\delta_{IJ}$. The simple positive roots are
$\alpha_I=\epsilon_I-\epsilon_{I+1}$ for $0\leq I\leq n-2$ and
$\alpha_{n-1}=2\epsilon_{n-1}$.
The half of the sum of the positive roots
is $\rho=n\epsilon_0+(n-1)\epsilon_1+\dots+\epsilon_{n-1}$.
We note also that $E_I$, $H_I$, and $F_I$ with $0\leq I\leq n-2$ form
a Chevalley
basis of the $sl(n)$ subalgebra in $sp(2n)$.

We describe in details the structure of the $so(d)$--$sp(2n)$ bimodule
$\cP_{n}^d(a)$
in the case $n=[d/2]$ and give several notes in the case $n>[d/2]$.
We suppose $n=[d/2]$ and let $\Lambda$ denote
the space of vectors $\sigma=(s_{n-1}, s_{n-2}, \cdots, s_0)$
with integer components satisfying $s_{n-1}\geq s_{n-2}\geq \cdots\geq
s_0\geq0$.
Thus $\sigma$ defines a highest weight of $so(d)$ and $\Lambda$ is the
space of dominant
highest weights corresponding to tensor modules. We also define a mapping
$\theta$ from $\Lambda$ to the space of $sp(2n)$ highest weights
\be\label{theta-map}
\theta(\sigma)=-\sum_{I=0}^{n-1}(s_I+\frac{d}{2})\epsilon_I.
\ee
The highest weight $\theta(\sigma)$ can be at the same time considered a $sl(n)$
highest weight.
By Howe duality $so(d)$ and $sp(2n)$ algebras mutually centralize each other
in $\cP_{n}^d(a)$, (\ref{pol}). The $so(d)$--$sp(2n)$ bimodule $\cP_{n}^d(a)$
has the structure~\cite{AdBarb}
\be\label{Pnd2}
 \cP_{n}^d(a) = \mathop{\oplus}\limits_{\sigma\in\Lambda}
       (V_\sigma \otimes U_{\theta(\sigma)})\;,
\ee
where $V_\sigma$ and $U_{\theta(\sigma)}$ are irreducible $so(d)$ and $sp(2n)$
modules with highest weights $\sigma$ and $\theta(\sigma)$ respectively.
The module $U_{\theta(\sigma)}$ is the generalized Verma module
induced from the finite dimensional irreducible $sl(n)$ module
$W_{\theta(\sigma)}$
with integer dominant $sl(n)$ highest weight $\theta(\sigma)$.
In other words, it means that $U_{\theta(\sigma)}$ is \textit{freely}
generated by
generators $T_{IJ}$ from $sl(n)$ module $W_{\theta(\sigma)}$.
The check that the $sp(2n)$ generalized Verma module $U_{\theta(\sigma)}$
with the highest weight
$\theta(\sigma)$ (\ref{theta-map}) is simple is reduced to the
standard application
of the Kac--Kazhdan criterion \cite{mazorchuk-generalized} that image of
$\theta(\sigma)+\rho$ under a reflection from the Weyl group can not belong to
the lattice of weights of~$U_{\theta(\sigma)}$.

In particular, the module $U_{\theta(\sigma)}$ is cofree with respect to
generators $T^{IJ}$ and therefore the cohomology of the operator $\Delta_{IJ}=C_{IJ}T^{IJ}$
in $U_{\theta(\sigma)}$ are
\begin{equation}
H^n(\Delta_{IJ})=\left\{\begin{array}{ll}
                          W_{\theta(\sigma)},&\qquad n=0,\\
                           0,&\qquad n>0.
                         \end{array}
\right.
\end{equation}

In the case $n>[d/2]$ the module $U_{\theta(\sigma)}$ is not isomorphic to
a generalized Verma module but is a quotient of a generalized Verma module.
In other words there are some relations between generators $T_{IJ}$.
Whenever, $n>[d/2]$ the decomposition (\ref{Pnd}) is the same but
the mapping $\theta$ is defined as follows.
Let $\sigma=(s_{n-1}, s_{n-2}, \cdots, s_{n-[d/2]})$
with integer components satisfying $s_{n-1}\geq s_{n-2}\geq \cdots\geq
s_{n-[d/2]}\geq0$
be a dominant integer $so(d)$ highest weight. We set
$s_0=s_1=\dots=s_{n-[d/2]-1}=0$.
Then, the mapping $\theta$ is given by (\ref{theta-map}). In this case
the generalized Verma module induced from the finite-dimensional irreducible
$sl(n)$ module $W_{\theta(\sigma)}$ contains singular vectors and its structure
for large $n-[d/2]$ is quite complicated.

\section{Homological reduction}
\label{sec:C}

Here we reproduce the proposition on the homological reduction  proved in \cite{\BGST,Barnich:2006pc}.
Let $\cH$ be a vector (super)space. Consider a bundle $\bundle \cH=\manX\times\cH\rightarrow \manX$,
where $\manX$ is a space-time manifold  with local coordinates $x^a$, and denote by $\Gamma(\bundle\cH)$
the space of sections of $\bundle\cH$. There are two gradings, the Grassmann parity and ghost number
defined on $\cH$ which are naturally extended to $\cH$-valued sections.
The BRST operator $\brst: \Gamma(\bundle\cH) \rightarrow\Gamma(\bundle\cH)$ is a Grassmann odd
differential of finite order in $x$-derivatives with coefficients in linear operators
in $\cH$.

We assume that an additional grading in $\cH$ can be introduced such that each graded
component is finite-dimensional.

\begin{prop}\label{prop:red}
Suppose $\cH$ to be equipped with an additional grading besides the
ghost number,
\begin{equation}
  \cH=\bigoplus_{i\geq 0} \cH_{i},\qquad \deg(\cH_{i})=i,
\end{equation}
and let the BRST operator $\brst$ have the form
\begin{equation} \label{eq:2diffeq}
  \brst= \brst_{-1}
  +\brst_0
  +\sum_{i\geq1}\brst_i,\qquad \deg(\brst_i)=i,
\end{equation}
with $\brst_i:\Gamma({\bundle\cH})_{j}\to\Gamma({\bundle\cH})_{i+j}$.
If $\brst_{-1}$ is independent of $x$ and contains no $x$-derivatives
then the cohomology $H(\brst_{-1},\Gamma(\bundle\cH))\simeq
\Gamma(\bundle\cE)$ for some vector bundle $\bundle\cE\subset\bundle\cH$ and the system
$(\brst,\Gamma({\bundle\cH}))$ can be consistently reduced to
$(\tilde\brst,\Gamma({\bundle\cE}))$, where the operator $\tilde\brst$
is the differential induced by $\brst$ in the cohomology of $\brst_{-1}$.
\end{prop}

Let us note that operator $\brst_{-1}$ acting on $\cH$ induces a
triple  decomposition $\bundle\cH = \bundle\cE\oplus \bundle\cF\oplus \bundle\cG$,
where $\Ker \brst_{-1}=\bundle\cE\oplus \bundle\cG$, $\bundle\cE\simeq
H(\brst_{-1},\bundle\cH)$, $\bundle\cG= {\rm Im}\,\brst_{-1}$, and
$\bundle\cF$ is a complementary subbundle. Then
$\st{\cG\cF}{\brst}$ is algebraically invertible and
$\tilde\brst$ is given by
\begin{equation}\label{tilde-brst}
  \tilde\brst
  =(\st{\bundle\cE\bundle\cE}{\brst}
  - \st{\bundle\cE\bundle\cF}{\brst}(\st{\bundle\cG\bundle\cF}{\brst})^{-1}
  \st{\bundle\cG\bundle\cE}{\brst})\;,\qquad\quad
\tilde\brst:\;\;\Gamma(\bundle{\cE})\to\Gamma(\bundle{\cE})\,.
\end{equation}
An explicit recursive construction for $\tilde\brst$ can be found in \cite{\BGST}.
Note also that if the
cohomology of $\brst_{-1}$ is concentrated in one degree then $\tilde
\brst=\brst_{0}$ considered as acting in $\Gamma(\bundle\cE)$.

In the case where equations of motion have the unfolded form $\brst \Phi_{(p)}=0$ with $\brst$
being a flat covariant differential acting on differential $p$-forms $\Phi_{(p)}$,
the respective  $\brst_{-1}$ was originally  identified as the $\sigma_{-}$-operator
\cite{Shaynkman:2000ts,Vasiliev:2001wa}.

\section{$Q$-cohomology for any $n$}
\label{sec:B}

The proof in the general case goes in exactly the same way as for $n=3$. Namely, one constructs operators ${\hat Q}_l$
associated to the upper-triangular subalgebras $U_l$ using the same rule as before, {\it i.e.} ghosts
$b_l$ enter the respective terms either as $\sd_l\dl{b_l}$ or as $\bsd{}^l b_l$.
For any $\hat Q_l$ and $\hat Q_{l+1}$ one finds that their difference is in the term $q_{l+1}$ replaced
by $\bar q_{l+1}$ that shows that the cohomology of all $\hat Q_{l}$ is identical. The difference
between $\hat Q_l$ and $\hat Q_{l+1}$ originates from the relation between the upper-triangular subalgebras
$U_l$ and $U_{l+1}$ that can be visualized as the exchange  $\sd_{l+1}\leftrightarrow\bsd{}^{l+1}$.

The representation space of operators $\hat Q_l$ is given by
\be
\phi = \phi^{(0)}+\phi^{(1)}+ \cdots \phi^{(n-1)}\equiv \sum_{k=0}^{n-1} b_{i_1} \cdots b_{i_k}\, \phi^{i_1 \cdots i_k}\;,
\ee
where $\phi^{i_1 \cdots i_k}$ are anti-symmetric tensors, and ${\rm gh}\,\phi^{(k)}=k$.

Let us start the analysis of $Q$-cohomology with operator $\hat Q_0$ that can be represented as
\begin{equation}
\hat Q_0=\sum_{i>j}\chi_j^i \,N_i{}^j+\sum_{i>j, j\neq 1} \chi^i_j\, b_i\dl{b_j}+\sum_{j\neq 1}\sd_j\dl{b_j}+
(\sd_1+\chi^j_1 b_j)\dl{b_1}\;,
\end{equation}
where $\chi^i_j\,\,\, i>j$ are ghosts associated to $N_i{}^j \,\,\, i>j$.
At the minimal ghost number $-(n-1)$ the cohomology of $\hat Q_0$ is
obviously given by
\be
\hat Q_0 \phi = 0\;,
\ee
where elements $\phi=\phi^{(n-1)}\equiv b_{1}\ldots b_{n-1}\phi_{n-1}$
are such that $\phi_{n-1}$ satisfies
\be
N_i{}^j \phi_{n-1} = 0\;\;\; i>j\;,
\quad
\sd_i \phi_{n-1} = 0\;,\quad  i=1,..., n-1\;,
\ee
{\it i.e.} the highest weight vectors for the upper-triangular subalgebra $U_{n-1}$.

Then one again finds that the
last term of $\hat Q_0$ can be treated as $q_1=(\sd_1+\chi^j_1 b_j)\dl{b_1}$ and consistently replaced with
the respective $\bar q_1$. The resulting operator is $\hat Q_1$.
Applying the same reasoning as in the case of $n=3$ one concludes that
the cohomology of operators $\hat Q_0$ and $\hat Q_1$ are isomorphic and the representatives
can be taken the same
\be
\hat Q_0 \phi = \hat Q_1\phi=0\;.
\ee
Solving these relations one gets
\be
\label{1out}
\dl{b_1}\phi^{(k)}=0\;,
\qquad k=0,1,..., n-2\;,
\ee
along with a set of linear combinations of Young symmetrizers applied to $\phi^{(k)}$. One
observes then that in the ghost number $n-2$  relation \eqref{1out} means that the only non-zero component
of $\phi^{(n-2)}$ is $b_2 ... b_{n-1}\phi^{23... n-1}\equiv b_2 ... b_{n-1}\phi_{n-2}$ which
satisfies
\be
N_i{}^j \phi_{n-2} = 0\;\;\; i>j\;,
\quad
\sd_i \phi_{n-2} = 0\;\; i=2,3,..., n-1\;,
\quad
\bsd{}^1\phi_{n-2} = 0\;,
\ee
{\it i.e.} the highest weight vectors for the upper-triangular subalgebra $U_{n-2}$.

Repeating the procedure for $q_2$ etc one finds $n$ operators $\hat Q_i$ such that
the cohomology representatives can be taken to satisfy
\be
\hat Q_m \phi = \hat Q_{m+1}\phi = 0\;,
\qquad m=0,..., n-1\;.
\ee
At the last step
of the above iterative procedure one is left with operator $\hat Q_{n-1}$ which
defines the cohomology in the maximal ghost degree 0 through the cocycle condition
\be
\hat Q_{n-1}\phi = 0\;.
\ee

This immediately gives the answer for the cohomology. Namely the representative  at ghost number
$-p, 0\leq p \leq n-1$ is given by
$b_{n-p} b_{n-p+1}\ldots b_{n-1}\phi_p$ with $\phi_p$ satisfying
\be
\ba{c}
N_i{}^j\phi_p=0\,\,\,\;\; i>j,
\\
\\
\label{classes}
 \sd_{n-p}\phi_p=\sd_{n-p+1}\phi_p=\ldots=\sd_{n-1}\phi_p=0\,,\qquad \bsd{}^1\phi_p=\ldots=\bsd{}^{n-p-1}\phi_p=0\,.
\ea
\ee
{\it i.e.} the highest weight vectors for the upper-triangular subalgebra $U_{p}$.

Summarizing the above one concludes that any cohomology class of the original BRST operator
$Q$ has a representative that can be chosen to  satisfy
\be
\hat Q_0 \phi = \hat Q_1 \phi = \ldots =\hat Q_{n-1}\phi = 0\;.
\ee
Recall that in addition $\phi_p$ satisfies $N_i{}^j\phi_p=0\,\,\, i>j,\;$ $T^{IJ}\phi_p=0$, and
$\hat\cN_i(b_{n-p}\ldots b_{n-1}\phi_p)=s_i(b_{n-p}\ldots b_{n-1}\phi_p)$.

\providecommand{\href}[2]{#2}\begingroup\raggedright\endgroup

%\bibliography{HSmaster}

\end{document}